\documentclass[12pt]{article}
\usepackage[english,activeacute]{babel}
\usepackage{natbib}
\usepackage{comment}
\usepackage{float}
\usepackage[hidelinks]{hyperref}
\usepackage{mathrsfs}
\usepackage{enumerate}
\usepackage[font={small,it}]{caption}
\usepackage{amsmath,amsfonts,amsthm,amssymb}
\usepackage{bm,rotating,multirow,dsfont,graphicx}
\usepackage[usenames, dvipsnames]{color}
\usepackage{url}
\usepackage{multicol}
\usepackage{multirow}
\usepackage[T1]{fontenc}
\usepackage{flafter}
\usepackage{appendix}
\usepackage{subfigure}
\usepackage{xcolor}
\makeatletter
\def\hlinewd#1{%
	\noalign{\ifnum0=`}\fi\hrule \@height #1 %
	\futurelet\reserved@a\@xhline}
\makeatother
\DeclareMathOperator*{\rank}{rank}
\DeclareMathOperator*{\logitinv}{logit^{-1}}
\DeclareMathOperator*{\logit}{logit}
\newcommand\iid{\mathrel{\overset{\makebox[0pt]{\mbox{\normalfont\tiny\sffamily iid}}}{\sim}}}
\newcommand\simiid{\mathrel{\overset{\makebox[0pt]{\mbox{\normalfont\tiny\sffamily iid}}}{\sim}}}
\newcommand\simind{\mathrel{\overset{\makebox[0pt]{\mbox{\normalfont\tiny\sffamily ind}}}{\sim}}}
\newcommand{\pr}[1]{\textsf{Pr}\left(#1\right)}

\newcommand\floor[1]{\lfloor#1\rfloor}
\newcommand{\ex}[1]{ \exp{ \left\{ #1 \right\} } }
\newcommand{\expec}[1]{\textsf{E}\left(#1\right)}

\newcommand{\diag}[1]{\text{diag}\left[#1\right]}
\def\@roman#1{\romannumeral #1}
\def\spacingset#1{\renewcommand{\baselinestretch}{#1}\small\normalsize}\spacingset{1}
\def\le{\left}
\def\ri{\right}
\def\del{\delta}

\def\be{\beta}
\def\lam{\lambda}
\def\ka{\kappa}
\def\ome{\omega}
\def\om{\omega}
\def\gam{\gamma}
\def\ga{\gamma}
\def\al{\alpha}
\def\eps{\epsilon}
\def\ups{\upsilon}
\def\sig{\sigma}
\def\si{\sigma}
\def\te{\theta}
\def\siv{\boldsymbol{\sigma}}
\def\kav{\boldsymbol{\kappa}}
\def\tauv{\boldsymbol{\tau}}
\def\lamv{\boldsymbol{\lambda}}
\def\omev{\boldsymbol{\omega}}
\def\phiv{\boldsymbol{\phi}}
\def\tev{\boldsymbol{\theta}}
\def\thev{\boldsymbol{\theta}}
\def\bev{\boldsymbol{\beta}}
\def\muv{\boldsymbol{\mu}}
\def\gamv{\boldsymbol{\gamma}}
\def\gav{\boldsymbol{\gamma}}
\def\alv{\boldsymbol{\alpha}}
\def\omv{\boldsymbol{\omega}}
\def\Ups{\mathbf{\Upsilon}}

\def\SIG{\mathbf{\Sigma}}
\def\Ga{\Gamma}
\def\mv{\boldsymbol{m}}
\def\tv{\boldsymbol{t}}
\def\wv{\boldsymbol{w}}
\def\sv{\boldsymbol{s}}
\def\xv{\boldsymbol{x}}
\def\yv{\boldsymbol{y}}
\def\zv{\boldsymbol{z}}
\def\a{\mathbf{a}}
\def\A{\mathbf{A}}
\def\B{\mathbf{B}}
\def\m{\mathbf{m}}
\def\F{\mathbf{F}}
\def\G{\mathbf{G}}
\def\R{\mathbf{R}}
\def\f{\mathbf{f}}
\def\C{\mathbf{C}}
\def\I{\mathbf{I}}
\def\Ima{\mathbf{I}}

\def\H{\mathbf{H}}
\def\V{\mathbf{V}}
\def\W{\mathbf{W}}
\def\Xm{\mathbf{X}}
\def\X{\mathbf{X}}
\def\Z{\mathbf{Z}}
\def\trans{\textsf{T}}
\def\one{\mathbf{1}}
\def\zerov{\boldsymbol{0}}
\def\onev{\boldsymbol{1}}
\def\mle{\text{mle}}

\begin{document}

\title{Some Case Studies Using \\ Bayesian Statistical Models}

\date{}
	
\author{
    Juan Sosa, Universidad Nacional, Colombia\footnote{jcsosam@unal.edu.co}\\
    Lina Buitrago, Universidad Nacional, Colombia\footnote{labuitragor@unal.edu.co }\\
}	 

\maketitle

\begin{abstract} 
We provide four case studies that use Bayesian machinery to making inductive reasoning. Our main motivation relies in offering several instances where the Bayesian approach to data analysis is exploited at its best to perform complex tasks, such as description, testing, estimation, and prediction. This work is not meant to be either a reference text or a survey in Bayesian statistical inference. Our goal is simply to provide several examples that use Bayesian methodology to solve data-driven problems. The topics we cover here, include problems in Bayesian nonparametrics, Bayesian analysis of times series, and Bayesian analysis of spatial data.
\end{abstract}

\noindent
{\it Keywords:} Bayesian Models. Gibbs Sampler. Hierarchical Models. Markov chain Monte Carlo. Statistical Inference.

\spacingset{1.1} % DON'T change the spacing!

\section{Introduction}

In this manuscript, we provide four case studies that use Bayesian machinery to making inductive reasoning. Our main motivation relies in offering several instances where the Bayesian approach to data analysis is exploited at its best to perform complex tasks, such as description (probabilistic summary of main data patterns), testing (evaluation of competing theories of data formation), estimation (evaluation of parameters in a presumed model), and prediction (forecast of missing or future observations).

We focus on model-based approaches to perform statistical inference. In particular, our case studies are totally based on Bayesian statistics. A statistical model fully describes the data generating process under which a given dataset might have arisen. Thus, the vector of observations $\yv = (y_1,\ldots,y_n)$ is assumed to be generated by an unknown probability distribution $P$ with probability density/mass function $p$. Therefore, we have that all information in the data is contained in $P$.

Furthermore, in Bayesian statistics we treat the model parameters that index $P$, $\tev=(\te_1,\ldots)\in\Theta$, as random variables, which in turn are assigned a prior distribution $\Pi$ with probability density/mass function $p$ (even though it is an abuse of mathematical notation, since the probability density/mass function associated with $P$ is also denoted with $p$). The prior distribution captures the uncertainty of the researcher about the value of $\tev$ before observing $\yv$ (another way to think about the prior is as summarizing all information about $\tev$ that is external to $\yv$).

If we use probability theory to express all forms of uncertainty associated with our statistical model $\mathcal{M}=(p(\yv\mid\tev),p(\tev))$, then Bayes theorem can base used to update our knowledge about $\tev$, through the posterior distribution
$$
p(\tev\mid\yv)\propto\frac{p(\yv\mid\tev)\,p(\tev)}{\int_\Theta p(\yv\mid\tev)\,p(\tev)\,\text{d}\tev}\,.
$$
The posterior combines information in the data with any other information external to it that has been encoded in the prior. Again, we slightly abuse notation by using $p$ to denote the distribution of $\yv\mid\tev$ as well as that of $\tev\mid\yv$.

The Bayesian approach is convenient for several reasons. First of all, in a Bayesian setting all forms of uncertainty quantification are granted. Moreover, it is natural to think about hierarchies, and therefore, borrowing of information (hierarchical models are used in non-Bayesian approaches, but they are not as natural). Also, with the advent of Markov chain Monte Carlo (MCMC) and other simulation-based approaches, computation has become in some ways easier than computation for non-Bayesian approaches (particularly for complex models). Lastly, since both observations and parameters are random variables, dealing with missing and censored data is simpler.

This work is not meant to be either a reference text or a survey in Bayesian statistical inference. Our goal is simply to provide several examples that use Bayesian methodology to solve data-driven problems. Therefore, we expect that the reader has a working understating of Bayesian methods for model building and computation. For a deep treatment of Bayesian data analysis methods, we refer the reader to \cite{gelman2014bayesian}, \cite{robert2007bayesian}, \cite{jackman2009bayesian}, among many others. Finally, the topics we cover here, include problems in Bayesian nonparametrics (e.g., \citealt{muller2015bayesian}), Bayesian analysis of times series (e.g., \citealt{prado-10}), and Bayesian analysis of spatial data (e.g., \citealt{banerjee2014hierarchical}).

\newpage

The rest of the document is structured as follows: Section \ref{sec_regression} presents an example using synthetic data in the context of nonparamteric regression; Section \ref{sec_time_series} offers the analysis of a times series using dynamic linear models; Section \ref{sec_spatial} provides an illustration of an space-time model for areal unit data; Section \ref{sec_species} exhibits a case regarding species sampling models; finally, Section \ref{sec_discussion} gives a brief discussion.

\section{Nonparametric regression modeling}\label{sec_regression}

Consider the Gaussian process (GP) regression setting (e.g., \citealt{taddy2010bayesian}) with a single continuous covariate,
\begin{equation}\label{eq_model}
y_i = f(x_i) + \epsilon_i\,,\qquad i=1,\ldots,n\,,
\end{equation}
where $y_i$ and $x_i$ are the response and covariate observations, respectively, $\epsilon_i$ are independent and identically distributed (iid) random errors from a $\textsf{N}(0, \sigma^2)$ distribution, and $f(\cdot)$ is the regression function, which is assigned a GP prior with constant mean function $\textsf{E}(f(x)\mid\mu) = \mu$, and power exponential covariance function $\textsf{Cov}(f(x), f(x') \mid \tau,\phi) = \tau^2\exp{ \left\{ -\phi | x - x' |^\alpha \right\} }$, with unknown $\tau > 0$ and $\phi > 0$, but fixed $0<\alpha\leq 2$ (the special cases for $\alpha = 1$ and $\alpha = 2$ correspond to the Exponential and Gaussian covariance functions, respectively). Here, $\alpha$ defines the smoothness of the function, $\tau^2$ determines the amplitude of oscillation, and $\phi$ delimits the volatility of orientation. Lastly, recall that based on the GP definition, for any collection of index points $t_1,\ldots,t_n$, the random vector $(x_{t_1},\ldots,x_{t_n})$ follows an $n$-variate Normal distribution with mean vector $(\mu(t_1),\ldots,\mu(t_n))$, and positive definite covariance matrix whose $(i,j)$-th element is given by $\textsf{Cov}(x_{t_i},x_{t_j}) = \tau^2\exp{ \left\{ -\phi | t_i - t_j |^\alpha \right\} }$.

In order to perform full Bayesian inference under this setting, we consider the following independent prior distributions for the remaining model parameters:
$$
\sigma^2\sim\textsf{IG}(a_{\sigma}, b_{\sigma})\,,\qquad \mu\sim\textsf{N}(a_\mu, b_\mu)\,,\qquad \tau^2\sim\textsf{IG}(a_\tau, b_\tau)\,,\qquad \phi\sim\textsf{U}(a_\phi, b_\phi)\,.
$$
%Figure \ref{fig_DAG_GPNR} provides a directed acyclic graph (DAG) representation of this model.
Note that by letting $\thev=(\theta_1,\ldots,\theta_n)$, with $\theta_i=f(x_i)$ for $i=1,\ldots,n$, denote the unknown function values, it follows that $\yv\mid\thev,\sig^2\sim\textsf{N}_n(\thev,\sig^2\I)$ and $\thev\mid\muv,\H\sim\textsf{N}_n(\muv,\H)$, with $\yv=(y_1,\ldots,y_n)$, $\muv = \mu\one$ and $\H = \tau^2\C$, where $\C = \left[\exp{ \left\{ -\phi | x_i - x_j |^\alpha \right\} }\right]$, which means that once $\thev$ has been integrated out $\yv\mid\muv,\sig^2,\H\sim\textsf{N}_n(\muv,\sig^2\I + \H)$.

%\begin{figure}[!h]
%    \centering
%    \includegraphics[scale=.85]{DAG_GPNR}
%    \caption{\footnotesize{DAG representation of a GP nonparametric regression model. Circles represent either random variables or random vectors, and the edges convey conditional independence. Squares represent fixed quantities (constants).}}\label{fig_DAG_GPNR}
%\end{figure}

%\begin{align*}
%p(\mathbf{\Upsilon}\mid\boldsymbol{y})&\propto (\sigma^2)^{-n/2}\exp\left\{ -\tfrac{1}{2\sigma^2} \textstyle\sum_{i=1}^n(y_i-\theta_i)^2 \right\} \times (\sigma^2)^{-(a_\sigma+1)} \exp\left\{ -b_\sig/\sigma^2 \right\} \\
%&\hspace{0.5cm}\times |\H|^{-1/2}\exp\left\{ -\tfrac12 (\thev-\muv)^{\textsf{T}} \H^{-1} (\thev-\muv) \right\} \times \exp\left\{ -\tfrac{1}{2b_\mu^2} (\mu-a_\mu)^2 \right\}\\
%&\hspace{0.5cm}\times (\tau^2)^{-(a_\tau+1)} \exp\left\{ -b_\tau/\tau^2 \right\}\,,
%\end{align*}
Now, we provide a Gibbs sampling algorithm designed to draw samples from the posterior distribution 
$$
p(\mathbf{\Upsilon}\mid\boldsymbol{y}) \propto p(\yv\mid\thev,\sig^2)\,p(\thev\mid\muv,\H)\,p(\mu)\,p(\sigma^2)\,(\tau^2)\,p(\phi)\,,
$$
where $\mathbf{\Upsilon} = (\thev, \sig^2, \mu, \tau^2, \phi)$. Let $\Ups^{(b)}$ be the vector of model parameters at iteration $b$ of the algorithm, for $b= 1,\ldots,B$. Given a starting point $\Ups^{(0)}$, we consider a Gibbs sampler (with a Metropolis step) updating $\Ups^{(b-1)}$ to $\Ups^{(b)}$ until convergence as follows:
\begin{enumerate}[1.]
  \item Sample $\thev\mid\sig^2,\mu,\tau^2,\phi,\yv\sim\textsf{N}_n(\m_{\thev},\V_{\thev})$, with 
  $$
  \V_{\thev} = \left(\sig^{-2}\I + \H^{-1}\right)^{-1}
  \qquad\text{and}\qquad
  \m_{\thev} = \V_{\thev}\left(\sig^{-2}\yv + \H^{-1}\muv\right)\,.
  $$
  \item Sample $\sig^2\mid\thev \sim \textsf{IG}\left(a_\sig + \frac{n}{2}, b_\sig + \frac12\sum_{i=1}^n (y_i - \theta_i)^2\right)$.
  \item Sample $\mu\mid \thev, \tau^2, \phi \sim \textsf{N}(m_\mu, V_\mu)$, with 
  $$
  V_\mu = (b_\mu^{-1} + \one^{\textsf{T}}\,\H^{-1}\one)^{-1}
  \qquad\text{and}\qquad
  m_\mu = V_\mu\left[a_\mu b_\mu^{-1} + \one^{\textsf{T}}\,\H^{-1}\thev\right]\,.
  $$
  \item Sample $\tau^2\mid\thev,\mu,\phi\sim \textsf{IG}\left(a_\tau + \frac{n}{2} , b_\tau + \frac12(\thev - \muv)^{\textsf{T}}\C^{-1}(\thev - \muv)\right)$.
  \item Sample $\phi$ using a Metropolis step:
  \begin{enumerate}[a.]
    \item Set $\eta=\text{logit}((\phi - a_\phi)/(b_\phi-a_\phi))$.
    \item Sample $\eta^*\sim\textsf{N}(\eta,\delta)$, with $\delta$ a tuning parameter.
    \item Compute $r=\exp{\left\{ \log p_\eta(\eta^*\mid \thev, \mu, \tau^2) - \log p_\eta(\eta\mid \thev, \mu, \tau^2) \right\}}$, with
        $$ 
        \log p_\eta(x\mid \thev, \mu, \tau^2) = \tfrac12\left( \log|\H^{-1}| - (\thev - \muv)^{\textsf{T}}\H^{-1}(\thev - \muv) \right) + x - 2\log(1+e^x) + k\,,
        $$
        where $k$ is a constant.
    \item Update $\phi$ to $\phi^*=(b_\phi-a_\phi)\text{expit}(\eta^*)+a_\phi$ with probability $r$.
  \end{enumerate}
\end{enumerate}

\begin{figure}[ht]
    \centering
    %\subfigure[Synthetic data]{\includegraphics[scale=.5]{data_GPNR}}
    \subfigure[Data and sample paths]  {\includegraphics[scale=.5]{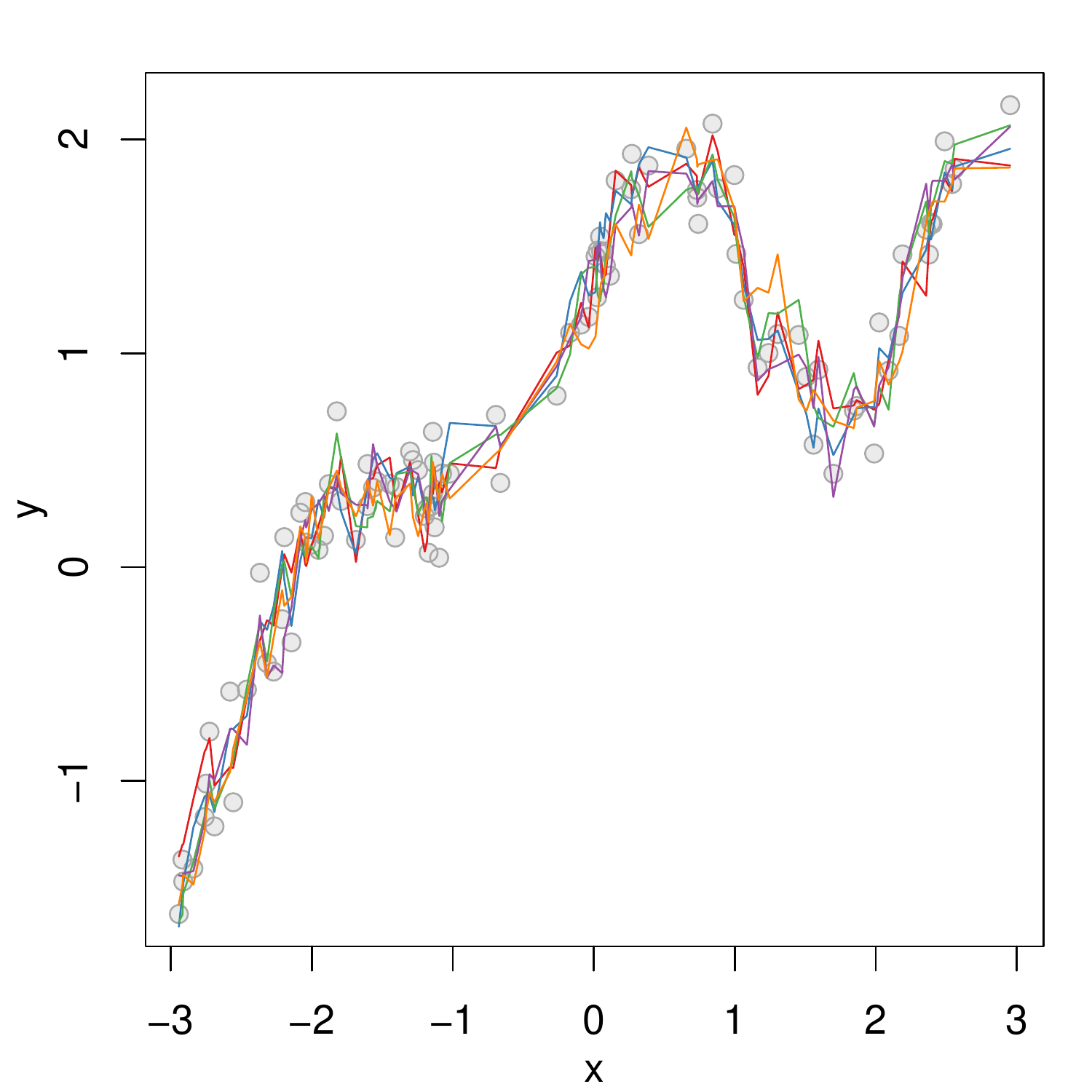}}
    \subfigure[Inference]     {\includegraphics[scale=.5]{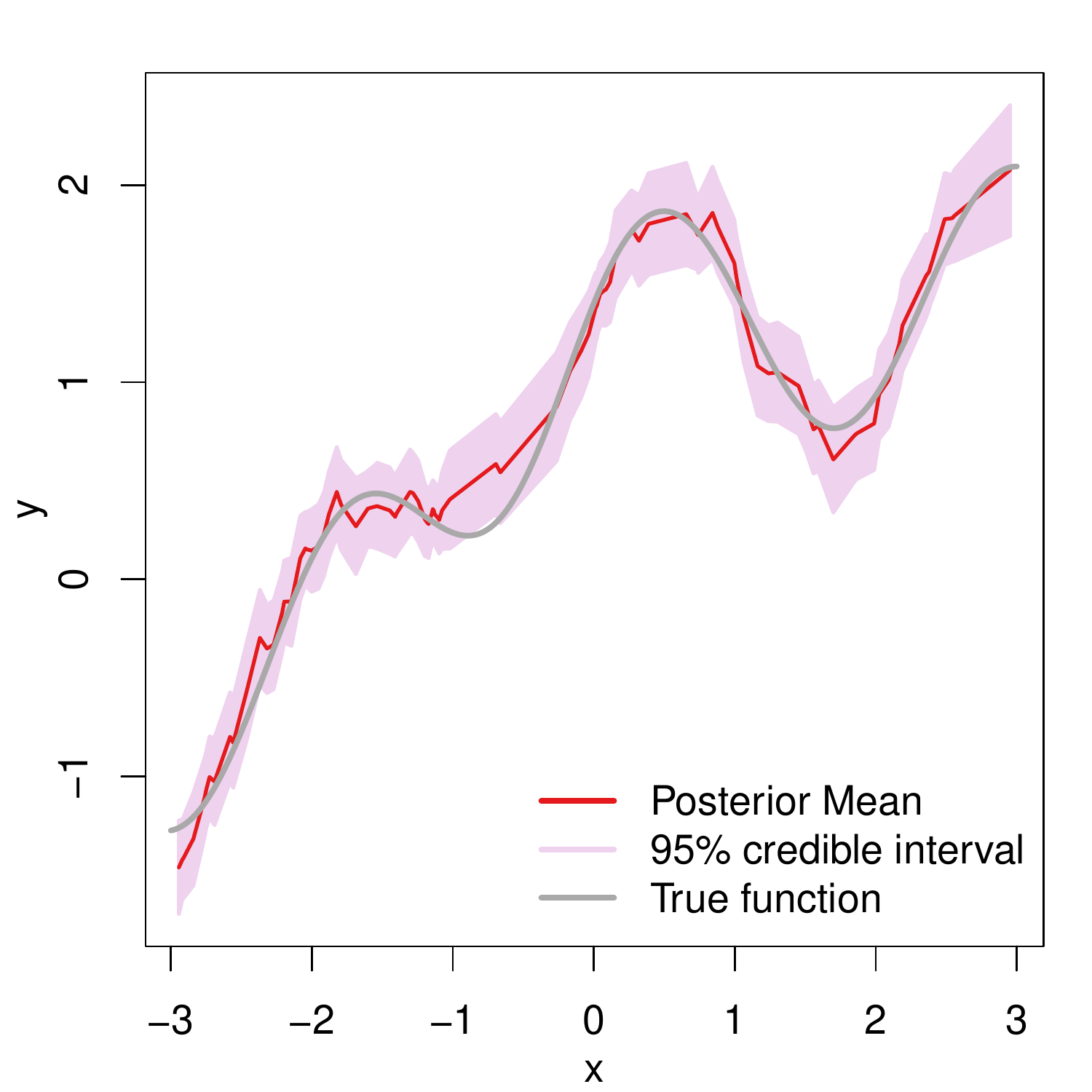}}
    \caption{\footnotesize{Data along with five sample paths chosen at random from the posterior distribution of $\thev$, and posterior inference on $f(\cdot)$.}}\label{fig_data_GPNR}
    %\caption{\footnotesize{(a) Synthetic data generated according to model \eqref{eq_model} with $\sigma=0.2$, $x_i\simiid\textsf{U}(-3,3)$, for $i=1,\ldots,100$, and $f(x)= 0.3 + 0.4x + 0.5\sin(2.7x) + \frac{1.1}{1 + x^2}$. (b) Five sample paths chosen at random from the posterior distribution of $\thev$. (c) Posterior inference on $f(\cdot)$.}}\label{fig_data_GPNR}
\end{figure}

Now, in order to test the previous algorithm, we generate synthetic data according to model \eqref{eq_model} with $\sigma=0.2$, $x_i\simiid\textsf{U}(-3,3)$, for $i=1,\ldots,100$, and the true regression function $f(x)= 0.3 + 0.4x + 0.5\sin(2.7x) + \frac{1.1}{1 + x^2}$ (this example is included in a technical report by Radford Neal, which is available on-line from \url{http://www.cs.toronto.edu/~radford/mc-gp.abstract.html}). Panel (a) in Figure \ref{fig_data_GPNR} shows the simulated dataset. 
Here, we adopt an empirical Bayes approach by setting $a_\mu=\bar{y}$, $a_\sig=a_\tau=2$, $b_\mu=b_\sig=b_\tau=s^2_y/3$, since marginally $\textsf{E}(y_i) = a_\mu = \bar{y}$ and $\textsf{Var}(y_i) = b_\mu+b_\sigma+b_\tau = s^2_y$; and also, $a_\phi = 0.1$ and $b_\phi=10$, which let $\phi$ vary across a wide range of values. Thus, we fit the model with $\alpha=1$ and run the algorithm choosing $\delta$ (step 5.) carefully to produce a reasonable average mixing rate between 30\% and 50\%.

The results presented below are based on $B = 25000$ samples from the posterior distribution $p(\mathbf{\Upsilon}\mid\boldsymbol{y})$ obtained after thinning the original chain every 10 observations and a burn-in period of $5000$ iterations. Chains mixed reasonably well and effective sample sizes for $\thev$ range from 22921 to 26436. Panel (a) in Figure \ref{fig_data_GPNR} shows five sample paths chosen at random from the posterior distribution of $\thev$. We see how these paths fit very well to the data. On the other hand, Table \ref{tab_summaries_GPNR} presents summarizes the posterior distribution of $\sig^2$, $\mu$, $\tau^2$, and $\phi$. Note that the 95\% quantile-based credible interval for $\sigma^2$ includes the true value (0.04). Finally, Panel (b) in Figure \ref{fig_data_GPNR} shows the posterior mean along with 95\% credible intervals for $f(\cdot)$. Almost the entire true function falls within the credible bands, which shows that the model specification as well as the prior elicitation was suitable in this case.

\begin{table}[ht]
\centering
\begin{tabular}{cccccc}
  \hline
 Parameter & Mean & SD & 2.5\% & 50\% & 97.5\% \\ 
  \hline
  $\sigma^2$ & 0.0334 & 0.0063 & 0.0228  & 0.0327 & 0.0476 \\ 
  $\mu$      & 0.6117 & 0.3932 & -0.1654 & 0.6144 & 1.3862 \\ 
  $\tau^2$   & 0.8075 & 0.4035 & 0.3303  & 0.7009 & 1.8700 \\ 
  $\phi$     & 0.3590 & 0.1740 & 0.1200  & 0.3296 & 0.7710 \\ 
   \hline
\end{tabular}
\caption{\footnotesize{Posterior summaries for $\sig^2$, $\mu$, $\tau^2$, and $\phi$.}}\label{tab_summaries_GPNR}
\end{table}

\section{Dynamic linear modeling}\label{sec_time_series}

\begin{figure}[ht]
    \centering
    \includegraphics[scale=.51]{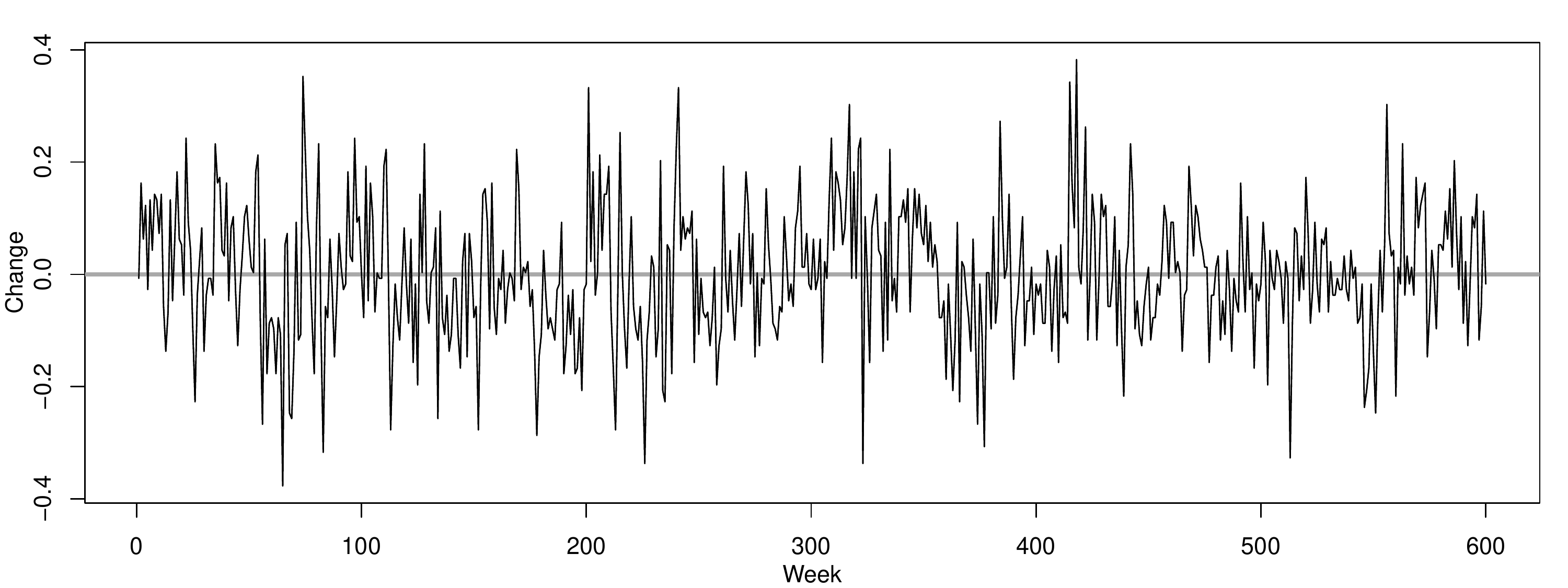}
    \caption{\footnotesize{Detrended zero-centered weekly change series of the U.S. 3-year Treasury Constant Maturity Interest Rate from March 18, 1988 to September 10, 1999.}}\label{fig_data_ts}
\end{figure}

Figure \ref{fig_data_ts} shows the detrended zero-centered weekly change series $y_1,\ldots,y_T$ associated with the U.S. 3-year Treasury Constant Maturity Interest Rate from March 18, 1988 to September 10, 1999. We have $T = 600$ equally spaced measurements in total. This series does not exhibit additional trends and seems to be stationary (which is confirmed below), so we do not differentiate the data again.

\subsection{Autoregressive model}

As a naive analysis of the series, we consider a model of the form
\begin{equation}\label{eq_AR}
y_t = \phi_1 y_{t-1} + \ldots + \phi_p y_{t-p} + \eps_t\,,\qquad \eps_t\mid v\simiid \textsf{N}(0, \ups)\,,
\end{equation}
which corresponds to an autoregressive model of order $p$, $\textsf{AR}(p)$. We fit this model to the data using a reference prior distribution of the form $p(\phiv,\ups)\propto 1/\ups$, with $\phiv=(\phi_1,\ldots,\phi_p)$, and the conditional likelihood
\begin{equation}\label{eq_cond_likelihood}
p(\yv\mid y_1,\ldots,y_p,\phiv,\ups) = \prod_{t=p+1}^T \textsf{N}(y_t\mid\f_t^{\textsf{T}}\phiv,\ups) = \textsf{N}_{T-p}(\yv\mid\F^{\textsf{T}}\phiv,\ups\I)
\end{equation}
where $\yv=(y_{p+1},\ldots,y_T)$ since the first $p$ observations are considered as initial values, $\f_t=(y_{t-1},\ldots,y_{t-p})$ for $t=p+1,\ldots,T$, and $\F=[\f_{p+1},\ldots,\f_{T}]$. This formulation corresponds to a regular linear model setting, and therefore, we can apply standard theory \cite[pp. 19-22]{prado-10}.

Combining the likelihood \eqref{eq_cond_likelihood} with the prior distribution $p(\phiv,\ups)$, we can easily obtain samples from the posterior distribution $p(\phiv,\ups\mid\yv)$ using direct sampling as follows: first, sample $\ups$ from $\textsf{IG}((n-p)/2,(n-p)\hat{\ups}^2/2)$, and then, for each $\ups$ in the previous step sample $\phiv$ from a $\textsf{N}_p(\hat{\phiv}_{\text{mle}}, \ups\,(\F\F^{\textsf{T}})^{-1})$. Using $p=3$, it follows that $n=T-p=597$, $(n-p)\hat{\ups}^2=7.6759$ is the residual sum of squares, which in turn leads to $\hat{v}_{\text{mle}}=0.0128$ the MLE of $v$, and $\hat{\phiv}_{\text{mle}}=(\F\F^{\textsf{T}})^{-1}\F\yv=(0.2283, 0.0018,0.1146)$ is the MLE of $\phiv$. Posterior summaries of 25000 samples from $p(\phiv,\ups\mid\yv)$ are provided in Table \ref{tab_summary_ar}. We see that the posterior parameter estimates are very similar to their MLE counterpart.

\begin{table}[ht]
\centering
\begin{tabular}{cccccc}
  \hline
   Parameter & Mean & SD & 2.5\% & 50\% & 97.5\% \\ 
  \hline
  $\phi_1$ & 0.2267 & 0.0407 & 0.1462  & 0.2267 & 0.3066 \\ 
  $\phi_2$ & 0.0063 & 0.0420 & -0.0753 & 0.0058 & 0.0896 \\ 
  $\phi_3$ & 0.1132 & 0.0408 & 0.0334  & 0.1132 & 0.1936 \\ 
  $\ups$   & 0.0130 & 0.0007 & 0.0116  & 0.0129 & 0.0145 \\ 
   \hline
\end{tabular}\caption{\footnotesize{Posterior summaries for $\phiv$ and $\ups$.}}\label{tab_summary_ar}
\end{table}

\subsection{Dynamic linear model}

Now, going forward in the analysis, we consider a popular model given in \cite[p. 561]{tsay2010analysis} for detecting outliers:
\begin{align}\label{eq_model_tsay}
    y_t &= \gam_t\al_t + x_t + \eps_t\,,\quad \gam_t\simiid \textsf{Ber}(0.2)\,,\quad \al_t\simiid \textsf{N}(0, 0.1)\,,\quad \eps_t\simiid \textsf{N}(0,0.01)\,, \\
    x_t &= \phi_1x_{t-1} + \ldots + \phi_px_{t-p} + \eta_t\,,\quad \eta_t\mid\omega\simiid \textsf{N}(0,\omega)\,.\nonumber
\end{align}
Under this formulation, $\gam_t$ denotes the presence or absence of an additive outlier at time $t$, and $\al_t$ its corresponding magnitude when present, and. Thus, additive outliers are allowed to occur at every time point with a probability equal to 0.2. Furthermore, $x_t$ defines an autoregressive process with order $p$. Then, we have $3T+p+1$ model parameters, namely, $\gamv=(\gam_1,\ldots,\gam_T)$, $\alv=(\al_1,\ldots,\al_T)$, $\xv=(x_1,\ldots,x_T)$, $\phiv=(\phi_1,\ldots,\phi_p)$, and $\omega$. Also, we also assume the conjugate priors $\phi_j\simiid \textsf{N}(a_\phi, b_\phi)$, for $j=1,\ldots,p$, and $\omega\sim\textsf{IG}(a_\omega, b_\omega)$.

Note that model \eqref{eq_model_tsay} is a non-Gaussian dynamic linear model (DLM). However, conditional on the outlier indicators $\ga_t$, this model can be written as a standard DLM. For instance, with $p = 3$ the model aquires the form:
\begin{align}\label{eq_CDLM}
    y_t &= \F^{\textsf{T}}\tev_t + \nu_t\,,        \qquad \nu_t\mid V_t \simind \textsf{N}(0,V_t)\,,\\
    \tev_t &= \G\tev_{t-1} + \omv_t\,, \qquad \omv_t\mid\W_t         \simind \textsf{N}_p(\boldsymbol{0}, \W_t)\,,\nonumber
\end{align}
with
$$
\F=(1,0,0)\,,\qquad\tev_t=(x_t,x_{t-1},x_{t-2})\,,\qquad V_t=v_{\ga_t} =\left\{
                                                                      \begin{array}{ll}
                                                                        0.01, & \hbox{$\ga_t=0$;} \\
                                                                        0.11, & \hbox{$\ga_t=1$,}
                                                                      \end{array}
                                                                    \right.
$$
and
$$
  \G=\begin{pmatrix}
       \phi_1 & \phi_2 & \phi_3 \\
       1 & 0 & 0 \\
       0 & 1 & 0 \\
     \end{pmatrix}\,,\qquad \omv_t=(\eta_t,0,0)\,,\qquad \W_t=\W=\begin{pmatrix}
                                                                 \omega & 0 & 0 \\
                                                                  0 & 0 & 0 \\
                                                                  0 & 0 & 0 \\
                                                                \end{pmatrix}\,.
$$
We use the previous fact to develop the second step of the MCMC algorithm described below. There, we implement a forward filtering backward sampling algorithm \cite[p. 137]{prado-10} to get samples from $p(\tev_{1:T}\mid\gav,\phiv,\omega)$, which automatically allow us to draw samples from $p(\xv \mid \gav, \phiv, \omega)$.

Now, we describe the general form of the MCMC to obtain samples from the posterior distribution 
$$
p(\Ups\mid\yv) \propto p(\yv \mid\gav,\alv,\xv)\,p(\xv\mid\phiv,w)\,p(\gav)\,p(\alv)\,p(\phiv)\, p(w)\,,
$$
where $\Ups = (\gav,\alv,\xv,\phiv,\omega)$ and $\yv = (y_1,\ldots,y_T)$. Full conditional distributions are derived looking at the dependencies in $p(\Ups\mid\yv)$. Let $\Ups^{(b)}$ be the vector of model parameters at iteration $b$ of the algorithm, for $b= 1,\ldots,B$. Given a starting point $\Ups^{(0)}$, we consider a Gibbs sampler (with a FFBS step) updating $\Ups^{(b-1)}$ to $\Ups^{(b)}$ until convergence as follows:
\begin{enumerate}
\item Sample $\al_t\mid\te_{t,1},\ga_t,y_t \sim \textsf{N}(m_t,V_t)$, for $t=1,\ldots,T$, with
      $$
      V_t=\frac1{10(10\ga_t+ 1)}
      \qquad\text{and}\qquad
      m_t = 100 V_t \ga_t(y_t-\te_{t,1})\,.
      $$
\item Sample $\ga_t\mid\te_{t,1},y_t \sim \textsf{Ber}(p_t)$, for $t=1,\ldots,T$, with
      $$
      \frac{p_t}{1-p_t}=\frac1{4\sqrt{11}}\,\exp\left\{ -\tfrac12(y_t - \te_{t,1})^2( 1/0.11 - 1/0.01 ) \right\}\,.
      $$
\item Sample $\phiv\mid\tev_{1:T},\omega\sim\textsf{N}_p(\m_{\phi},\V_{\phiv})$, with
      $$
        \V_{\phiv} = \left(b_\phi^{-1}\I + \frac{1}{\omega} \sum_{t=1}^T\tev_{t-1}\tev_{t-1}^{\textsf{T}}\right)^{-1}
        \quad\text{and}\quad
        \m_{\phi} = \V_{\phiv}\left(a_\phi b_\phi^{-1}\mathbf{1} + \frac{1}{\omega}\sum_{t=1}^T\te_{t,1}\tev_{t-1}\right)\,.
      $$
\item Sample $\omega\mid\tev_{1:T},\phiv\sim\textsf{IG}(A_\omega, B_\omega)$, with
      $$
        A_\omega = a_\omega + \frac{T}2
        \qquad\text{and}\qquad
        B_\omega = b_\omega + \frac12\sum_{t=1}^T(\te_{t,1}-\phiv^{\textsf{T}}\tev_{t})^2\,.
      $$
\item Sample $\tev_t$, for $t=1,\ldots,T$, using a FFBS step. Here we use standard DLM notation, in particular that corresponding to the Kalman filter \cite[Theorem 4.1, pp. 103-105]{west1999bayesian}:
\begin{enumerate}[a.]
    \item Use the filter equations to compute $\a_t$, $\R_t$, $\m_t$, and $\C_t$, for $t=1,\ldots,T$:
    $$
    \a_t=\G_t\m_{t-1}\,,\, \R_t = \G_t\C_{t-1}\G_t^{\textsf{T}} + \W_t\,,\, \m_t = \a_t+\A_t e_t\,,\,\C_t = \R_t-Q_t\A_t\A_t^{\textsf{T}}\,,
    $$
    with $e_t=y_t-f_t$, $f_t=\F_t^{\textsf{T}}\a_t$, and $Q_t = \F_t^{\textsf{T}}\R_t\F_t + V_t$.
    \item At time $t=T$, sample $\tev_T\sim \textsf{N}_p(\m_T, \C_T)$.
    \item For $t=(T-1),\ldots,0$, sample $\tev_t\sim \textsf{N}(\m_t^*, \C_t^*)$, where
          $$
             \m_t^*=\m_t+\B_t(\tev_{t+1} - \a_{t+1})
             \qquad\text{and}\qquad
             \C_t^*=\C_t - \B_t\R_{t+1}\B_t^{\textsf{T}}\,,
          $$
          with $\B_t=\C_t\G_{t+1}^{\textsf{T}}\R_{t+1}^{-1}$.
\end{enumerate}
\end{enumerate}

Now, we implement the MCMC provided above. To do so, we need to choose appropriate values for $a_{\phi}$, $b_\phi$, $a_\omega$, and $b_\omega$. 
First, $a_{\phi}=0$ and $b_\phi=0.25$ are reasonable values in accordance with our introductory analysis of the series using a standard autoregressive model.
On the other hand, we weakly concentrate the prior distribution of $\omega$ around $\hat\ups_{\text{mle}}=0.0128$ from our introductory analysis, by setting $a_\omega=2$ and $b_\omega=\hat\ups_{\text{mle}}$, which leads to $\textsf{E}(\omega)=\hat\ups_{\text{mle}}$ with infinite variance. The results presented below are based on $B = 25000$ samples from the posterior distribution $p(\mathbf{\Upsilon}\mid\boldsymbol{y})$ obtained after a burn-in period of $5000$ iterations. Chains mixed very well and there is no sign of lack of convergence. Effective sample sizes range from 398.7 to 29019.3.

\begin{table}
\centering
\begin{tabular}{cccccc}
  \hline
   Parameter & Mean & SD & 2.5\% & 50\% & 97.5\% \\ 
  \hline
  $\phi_1$ & 0.4692 & 0.1605 & 0.1618  & 0.4665 & 0.7904 \\  
  $\phi_2$ & 0.1341 & 0.1932 & -0.2694 & 0.1492 & 0.4727 \\ 
  $\phi_3$ & 0.1296 & 0.1367 & -0.1495 & 0.1309 & 0.4029 \\ 
  $\omega$ & 0.0017 & 0.0004 & 0.0010  & 0.0016 & 0.0026 \\ 
   \hline
\end{tabular}\caption{\footnotesize{Posterior summaries for $\phiv$ and $\omega$.}}\label{tab_summary_dlm}
\end{table}

\begin{figure}[t]
      \centering
      \subfigure[$\textsf{Pr}(\gamma_t=1\mid\yv)$]{\includegraphics[scale=.5]{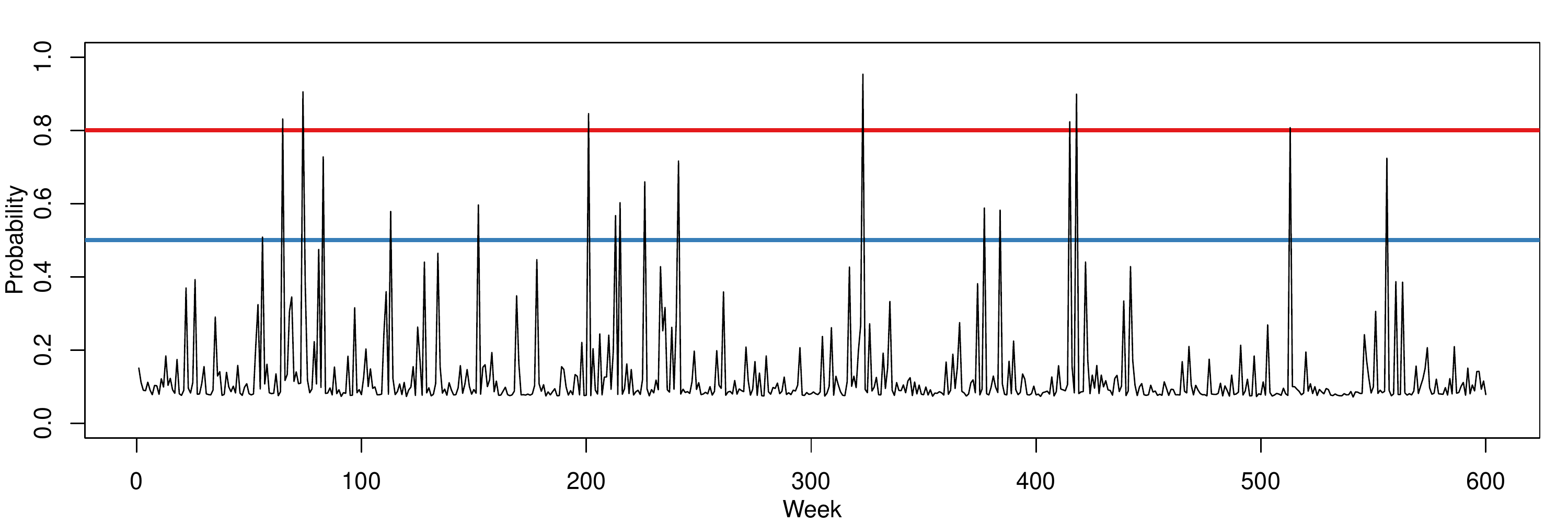}}
      \subfigure[$\textsf{E}(\alpha_t\mid\yv)$]{\includegraphics[scale=.5]{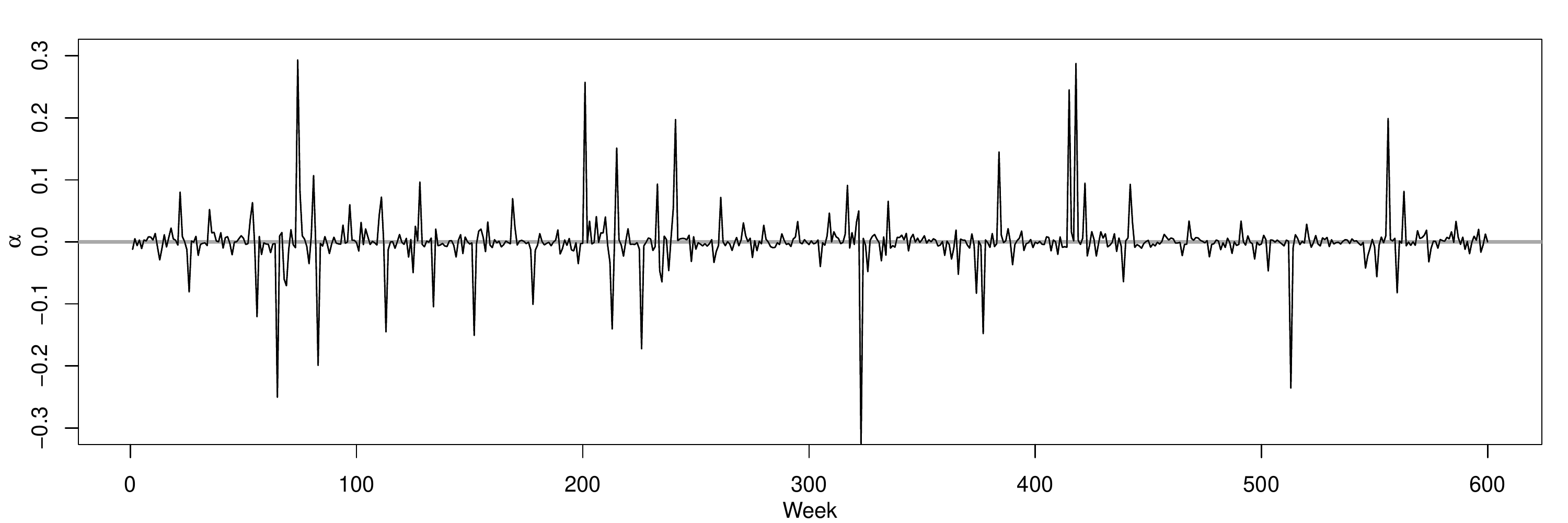}}
      \caption{\footnotesize{(a) Posterior probability that an observation is an outlier (lines correspond to 0.5 and 0.8). (b) Posterior mean of an outlier size.}}\label{fig_post_gamma_alpha}
\end{figure}

\begin{figure}[ht]
      \centering
      \subfigure[$\textsf{E}(x_t\mid\yv)$]{\includegraphics[scale=.5]{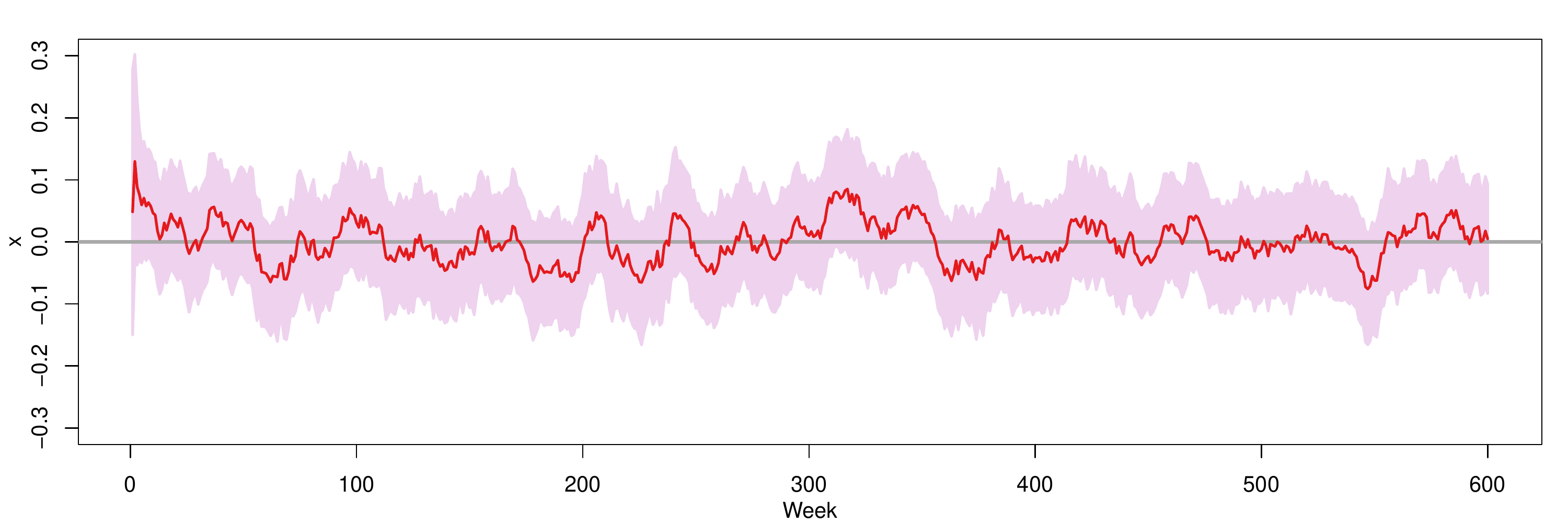}}
      \subfigure[$\textsf{E}(\gamma_t\alpha_t + x_t\mid\yv)$]{\includegraphics[scale=.5]{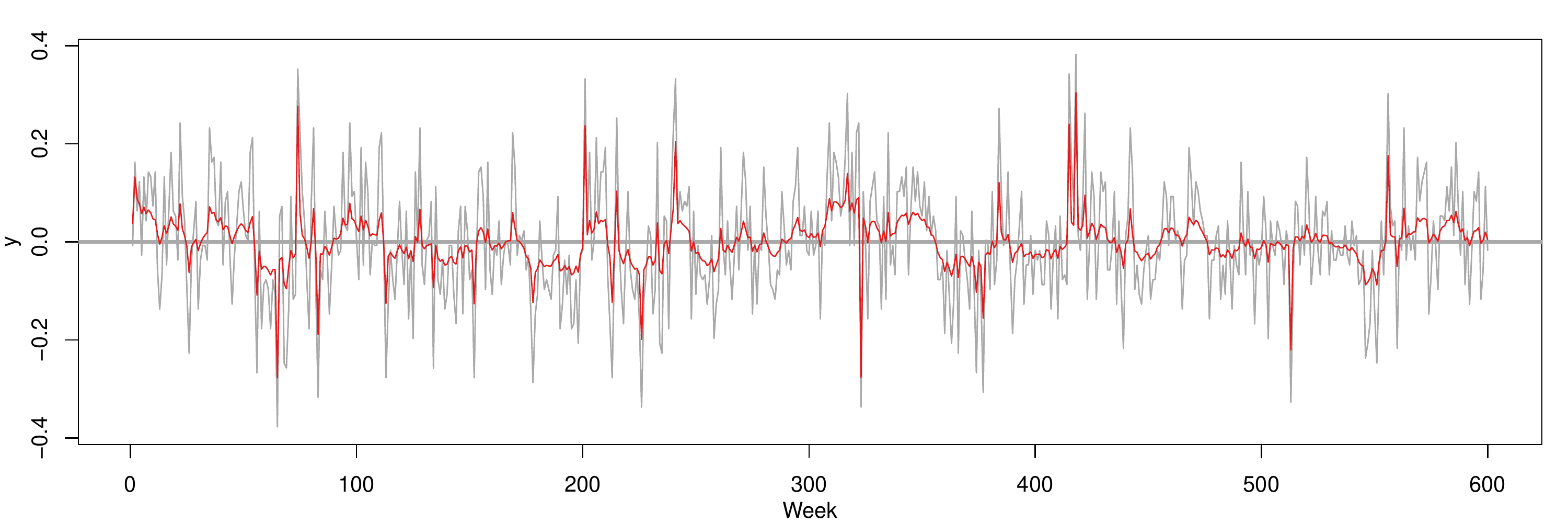}}
      \caption{\footnotesize{(a) Posterior inference on the autoregressive process. (b) Data (gray) and fitted values (red).}}\label{fig_post_x_y}
\end{figure}

Table \ref{tab_summary_dlm} display summaries for the posterior distribution of $\phiv$ and $\omega$. Estimates for $\phi_3$ and $\omega$ are consistent with those obtained in Table \ref{tab_summary_ar} using a standard autoregressive model. However, estimates for $\phi_1$ and $\phi_2$ are utterly different. Such a difference is expected since the autoregressive model does not consider additive outliers, and as a consequence, is not able to characterize fundamental features of the series such as its mean evolution. On the other hand, Figure \ref{fig_post_gamma_alpha} shows the posterior probability that an observation is an outlier $\textsf{Pr}(\gamma_t=1\mid\yv)$ along with the corresponding posterior mean of an outlier size $\textsf{E}(\alpha_t\mid\yv)$, for $t=1,\ldots,T$. We identify 18 observations such that $\textsf{Pr}(\gamma_t=1\mid\yv)>0.5$, seven of which are greater than 0.8, namely, observation at weeks  65, 74, 201, 323, 415, 418, and 513. \citet[p. 564]{tsay2010analysis} also classifies these points as outliers (employing an approach with mild differences) and highlight observations at times $t=323$ (May 20, 1994) and $t=201$ (January 17, 1992). At the former, there was a 0.6\% drop in the weekly interest rate, while at the later, there was a jump of about 0.35\%. Finally, Figure \ref{fig_post_x_y} displays the estimated autoregressive process as well as the fitted values along with the original data. We see that the model captures reasonably well the ``volatility'' of the series.

\section{Areal unit space-time modeling}\label{sec_spatial}

We consider an ecological application in which a spatiotemporal model for areal unit data is considered for modeling a set of binary observations $y_{i,t}= y_t(\sv_i)$, for $i=1,\ldots,I$ and $t=1,\ldots,T$, in a discrete time period over a fixed lattice $\sv_1,\ldots,\sv_I$. Here, we consider the Google flu data corresponding to the first 12 weeks of 2013 for the contiguous continental states in the United States (Alaska and Hawaii are omitted from the analysis). Thus, we record a binary variable $y_{i,t}$ that is equal to 1 if the Google flue index for state $i$ during week $t$ is greater than 7500 cases, and 0 otherwise.

We have that the observations $y_{i,t}$ are Bernoulli distributed with probability of success $\pi_{i,t}$, i.e., $y_{i,t}\mid\pi_{i,t}\sim \textsf{Ber}(\pi_{i,t})$. 
Here, we consider the following areal hierarchical generalized linear model \cite[Sec. 6.4]{banerjee2014hierarchical} including a temporal component:
\begin{align}\label{eq_model_spatial}
    y_{i,t}\mid\ome_{i,t} &= \left\{
                   \begin{array}{ll}
                     1, & \hbox{if $\ome_{it}>0$;} \\
                     0, & \hbox{if $\ome_{it}\leq0$,}
                   \end{array}
                 \right.\\
    \ome_{i,t}\mid \bev, \gav,\xi, \te_{i,t}, \phi_{i,t}, \ka_t &\simind\textsf{N}\left(\xv_{i}^{\textsf{T}}\bev + \zv_{i}^{\textsf{T}}\gav + \xi t + \te_{i,t} + \phi_{i,t}, 1/\ka_{t} \right)\,,\nonumber
\end{align}
with a hierarchical prior distribution of the form
\begin{align}\label{eq_prior_spatial}
    \be_k    &\simiid \textsf{N}(0,\si^2_0)\,,&
    \ga_\ell &\simiid \textsf{CAR}(1/\si^2_0)\,,&
    \xi      &\sim    \textsf{N}(0,\si^2_0)\,,&\\ \nonumber
    \te_{i,t}\mid\tau_t  &\simind \textsf{N}(0,1/\tau_{t})\,,&
    \phi_{i,t}\mid\lam_t &\simind \textsf{CAR}(\lam_t)\,,&
    \ka_t    &\simiid     \textsf{G}(\nu_0/2,\nu_0/2)\,,&\\ 
    \tau_t   &\simiid \textsf{G}(a_\tau,b_\tau)\,,&
    \lam_t   &\simiid \textsf{G}(a_\lam,b_\lam)\,,& \nonumber
\end{align}
for $k=1\ldots,K$, $\ell=1,\ldots,L$, where $\nu_0$ is a fixed positive integer, $\si^2_0,a_\tau, b_{\tau},a_{\lam},b_{\lam}$ are fixed positive real numbers, and $\bev=(\be_1,\ldots,\be_K)$, $\xv_{i}=\xv(\sv_i)=(x_{1}(\sv_i),\ldots,x_{K}(\sv_i))$, $\gav=(\ga_1,\ldots,\ga_L)$, and $\zv_{i}=\zv_{t}(\sv_i)=(z_{1}(\sv_i),\ldots,z_{L}(\sv_i))$.

The link function $g(\cdot)$ relating the $\pi_{i,t}$ and the linear predictor is based in a Student-$\textsf{t}$ distribution with $\nu_0$ degrees of freedom, which is particularly useful to detect rare events (those with small $\pi_{i,t}$). Since $\ome_{i,t} = \eta_{i,t} + \phi_{i,t} + \eps_{i,t}$ with $\eta_{i,t}=\xv_{i}^{\textsf{T}}\bev + \zv_{i}^{\textsf{T}}\gav + \xi t + \te_{i,t}$, $\eps_{i,t}\sim \textsf{N}(0, 1/\ka_{t})$ and $\ka_t\sim \textsf{G}(\nu_0/2,\nu_0/2)$, then the marginal distribution of $\eps_{i,t}$ (and therefore $\ome_{it}$) is Student-$\textsf{t}$ with $\nu_0$ degrees of freedom. Thus, due to the symmetry of the $\textsf{t}$ distribution, we have that $\pr{y_{i,t}=1\mid\ome_{i,t}}=\pr{\eps_{it} \leq \eta_{i,t}}$, and therefore, $\pi_{i,t}=F_{\nu_0}(\eta_{i,t})$, where $F_{\nu_0}$ is the cumulative distribution functions of a random variable following a $\textsf{t}$ distribution with $\nu_0$ degrees of freedom, which means that the link function is $g= F_{\nu_0}$.

Note that the main effects for non-spatiotemporal covariates have a standard linear regression structure $\xv_{i,t}^{\textsf{T}}\bev$, with a weakly informative prior for $\bev$ \citep[p. 55]{gelman2014bayesian}. Also, the main effects for space covariates have a standard linear regression structure $\zv_{i,t}^{\textsf{T}}\gav$, following an improper conditionally autoregressive (CAR) model for $\gav$ \citep[p. 81]{banerjee2014hierarchical}. Furthermore, the main effects for time are included in the linear predictor through $\delta_t=\xi t$, with a weakly informative prior for $\xi$. Lastly, the main effects for spatiotemporal interactions $\theta_{i,t}+\phi_{i,t}$ have two components. On the one hand, $\theta_{i,t}$ captures region-wide heterogeneity (unstructured variation) by means of $\textsf{N}(0,1/\tau_t)$, where $\tau_t$ is a precision parameter controlling the magnitude of $\te_{i,t}$. And on the other, $\phi_{i,t}$ captures regional clustering (spatially structured variability) via $\textsf{CAR}(\lam_{t})$, where $\lam_t$ is a precision parameter as before.

CAR models are very convenient computationally, since the method for exploring the posterior distribution is in itself a conditional algorithm, the Gibbs sampler, which also eliminates the need for matrix inversion. However, CAR models have two main theoretical and computational challenges, namely, model impropriety and precision parameter selection. \citet[p. 155]{banerjee2014hierarchical} recommend to ignore impropriety and work with the intrinsic CAR specification, since we are using it just as a prior distribution and not to model the data directly. Even though this is the usual approach it requires some care: This improper CAR prior is a pairwise difference prior that is identified only up to an additive constant. Thus, in order to identify an intercept term (say $\be_1$), we must add the constraint $\sum_{i=1}^I \phi_{i,t}=0$ (this constrain is imposed numerically during computation). Moreover, $\tau_t$ and $\lam_t$ cannot be chosen arbitrarily large, since $\te_{i,t}$ and $\phi_{i,t}$ would become unidentifiable. After all, we have only a single $y_{i,t}$ in order to estimate two random effects for each $i$ and each $t$. To make the prior specification $\tau_t \sim \textsf{G}(a_\tau,b_\tau)$ and $\lam_t \sim \textsf{G}(a_\lam,b_\lam)$ sensible, \citet[p. 156]{banerjee2014hierarchical} recommend a prior specification that leads to
\begin{equation}\label{eq_prior_elicitation}
    \textsf{SD}(\te_{i,t})=\frac1{\sqrt{\tau_t}}\approx\frac1{0.7\sqrt{\bar{m}\,\lam_t}}\approx \textsf{SD}(\phi_{i,t})\,,
\end{equation}
where $\bar{m}$ is the average number of neighbors.

\begin{figure}[t]
    \centering
    \subfigure[States neighbor matrix]  {\includegraphics[scale=.25]{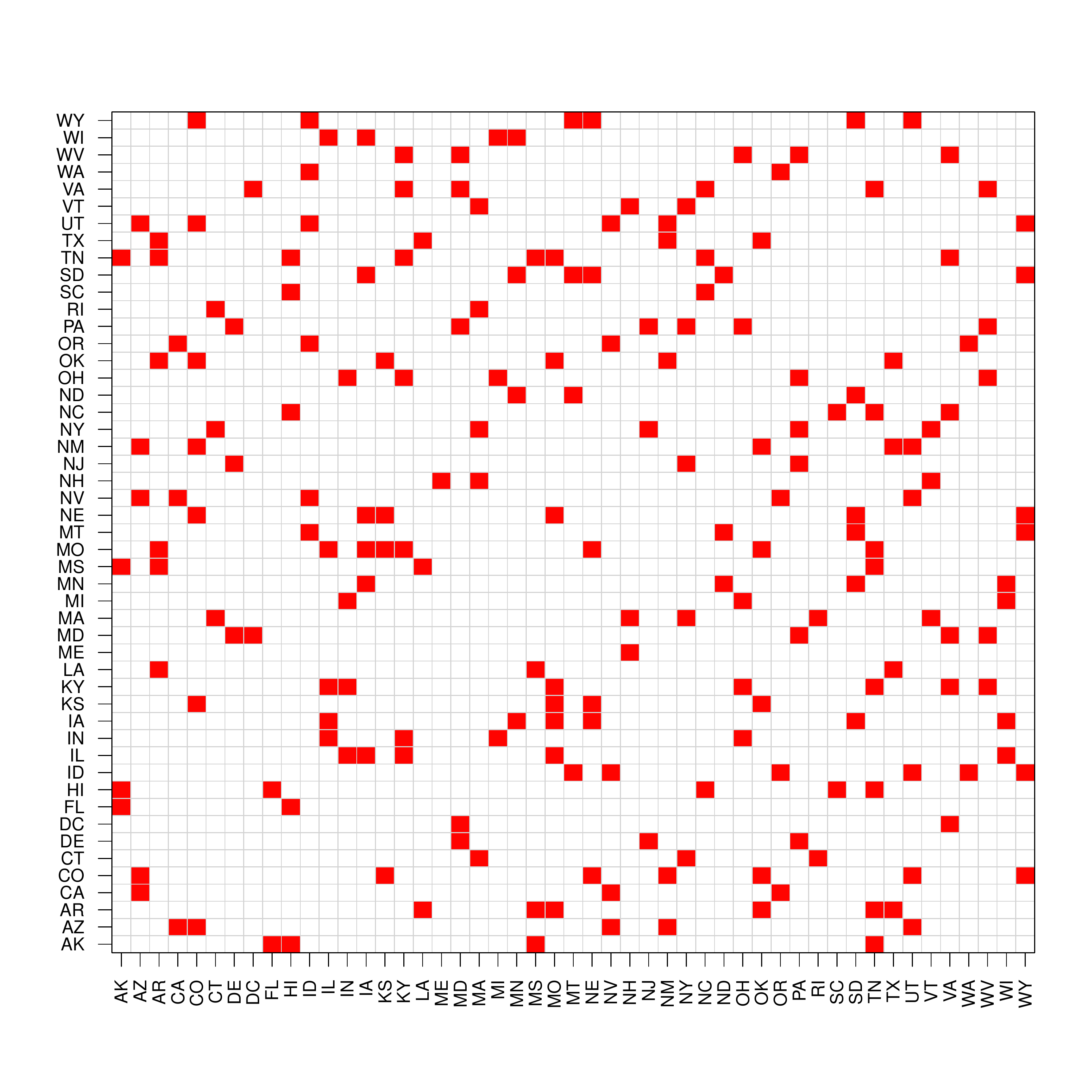}}
    \subfigure[Regions neighbor matrix] {\includegraphics[scale=.25]{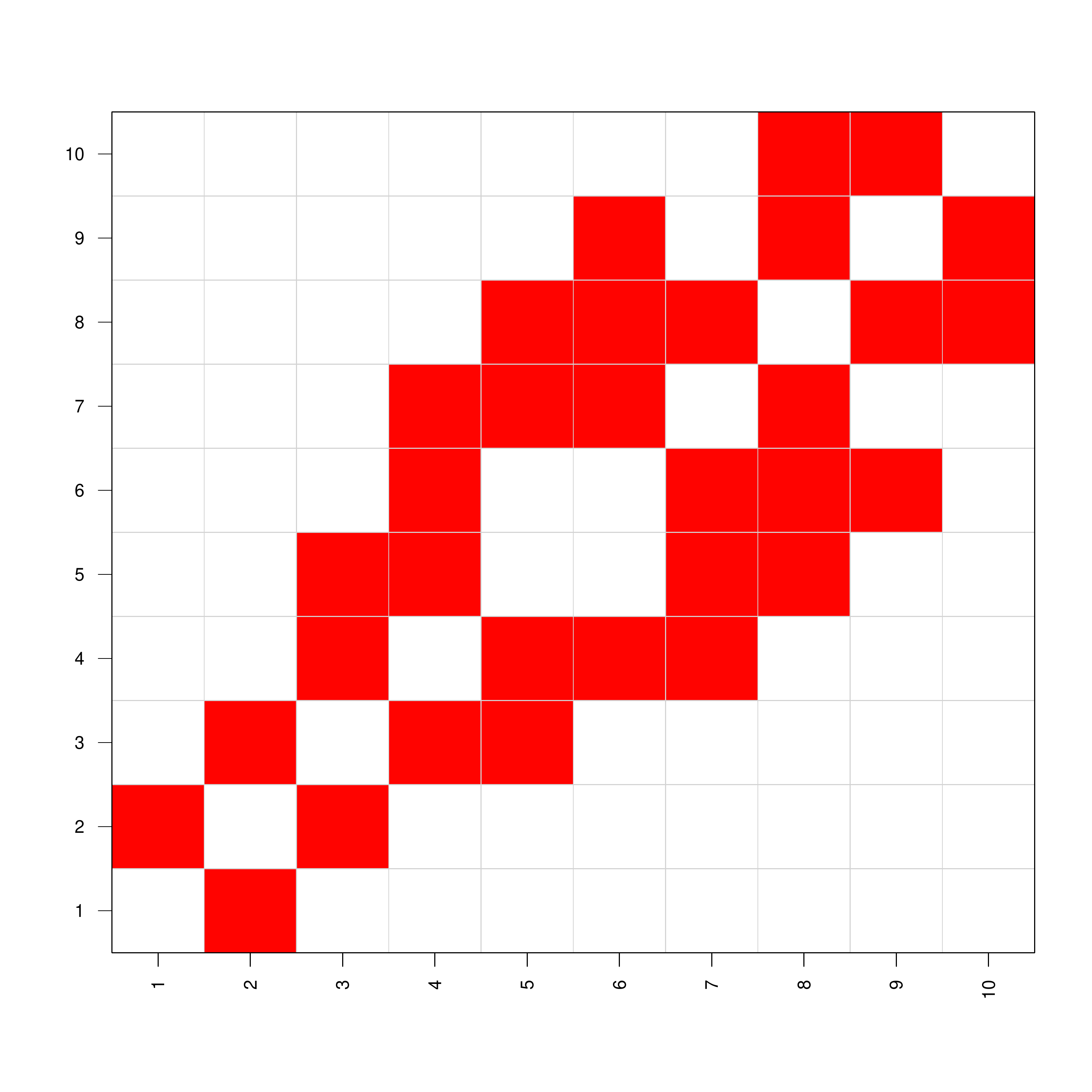}}
    \caption{\footnotesize{States and regions neighbors used to specify precision matrices. Red boxes denote neighbor elements.}}\label{fig_adj_mat}
\end{figure}

Define $\yv=(y_{1,1},\ldots,y_{I,T})$, $\omev = (\omega_{1,1},\ldots,\omega_{I,T})$, $\thev = (\theta_{1,1},\ldots,\theta_{I,T})$, $\phiv = (\phi_{1,1},\ldots,\phi_{I,T})$, $\kav = (\ka_1,\ldots,\ka_T)$, $\tauv = (\tau_1,\ldots,\tau_T)$, $\lamv = (\lam_1,\ldots,\lam_T)$, and also, $\Ups = (\omev,\tev,\phiv, \bev, \gav, \xi)$. Parameter estimates can be obtained from the posterior distribution
$$
p(\Ups\mid\yv) \propto p(\yv\mid\omev)\,p(\omev\mid\tev,\phiv,\bev, \gav, \xi, \kav)\,p(\tev\mid\tauv)\,p(\phiv\mid\lamv)\,p(\bev)\,p(\gav)\,p(\xi)\,p(\kav)\,p(\tauv)\,p(\lamv)\,,
$$
where according to \eqref{eq_model_spatial} and \eqref{eq_prior_spatial},
\begin{align*}
p(\omev\mid\bev,\gav,\tev,\xi,\phiv,\kav)&=\prod_{i=1}^I\prod_{t=1}^T \textsf{N}\left(\ome_{i,t}\mid\xv_{i}^{\textsf{T}}\bev + \zv_{i}^{\textsf{T}}\gav + \xi t + \te_{i,t} + \phi_{i,t}, 1/\ka_{t} \right)\,,\\
p(\tev\mid\tauv) &= \textsf{N}_{IT}\left(\tev\mid\boldsymbol{0}, \textsf{diag}[(1/\tau_{1})\I_I,\ldots,(1/\tau_{T})\I_I]\right),\\
p(\phiv\mid\lamv) &= \textsf{CAR}\left(\phiv\mid\textsf{diag}[\lam_{1} \W,\ldots,\lam_{T} \W] \right)\,,\\
p(\bev)\,p(\gav)\,p(\xi)\,p(\kav)\,p(\tauv)\,p(\lamv)&= p(\xi)\times \prod_{k=1}^K p(\be_k) \times\prod_{\ell=1}^L p(\ga_\ell)\times \prod_{t=1}^T\,p(\ka_t)\,p(\tau_t)\,p(\lam_t)\,,
\end{align*}
with $p(\xi), p(\be_k), p(\ga_\ell), p(\ka_t), p(\tau_t), p(\lam_t)$ specified as in \eqref{eq_prior_spatial}, and $\W=[W_{i,j}]$ is a square matrix of size $I\times I$ such that
$$
W_{i,j}=\left\{
\begin{array}{ll}
m_i,       & \hbox{$i=j$;} \\
-1,        & \hbox{$i\sim j$;} \\
\,\,\,\,\,0, & \hbox{otherwise,}
\end{array}
\right.
$$
where $m_i$ is the number of neighbors of $\sv_i$ (see Figure \ref{fig_adj_mat}), and $i\sim j$ means that units $i$ and $j$ are neighbors. Full conditional distributions are derived looking at the dependencies in $p(\Ups\mid\yv)$. Let $\Ups^{(b)}$ be the vector of model parameters at iteration $b$ of the algorithm, for $b= 1,\ldots,B$. Given a starting point $\Ups^{(0)}$, we consider a Gibbs sampler (with a step with a numerical constraint) updating $\Ups^{(b-1)}$ to $\Ups^{(b)}$ until convergence as follows:
\begin{enumerate}[1.]
\item Sample
$$
\ome_{i,t}\mid\text{rest} \sim
\left\{
  \begin{array}{ll}
    \textsf{TN}_{(-\infty,0]}\left( \xv_{i}^{\textsf{T}}\bev + \zv_{i}^{\textsf{T}}\gav + \xi t + \te_{i,t} + \phi_{i,t}, 1/\ka_{t} \right), & \hbox{$y_{it}=0$;} \\
    \textsf{TN}_{(0,+\infty)} \left( \xv_{i}^{\textsf{T}}\bev + \zv_{i}^{\textsf{T}}\gav + \xi t + \te_{i,t} + \phi_{i,t}, 1/\ka_{t} \right), & \hbox{$y_{it}=1$,}
  \end{array}
\right.
$$
for $i=1,\ldots,I$, $t=1,\ldots,T$.
\item Sample 
$$
\bev\mid\text{rest}\sim \textsf{N}_K\left( \V_{\bev}\le[\X^{\textsf{T}}\SIG^{-1}_{\omev}(\wv-\Z\gav-\tv\xi-\tev-\phiv)\ri], \V_{\bev} \right)
$$
where $\V_{\bev}=(\si_0^{-2}\I_K+\X^{\textsf{T}}\SIG^{-1}_{\omev}\X)^{-1}$, $\SIG^{-1}_{\omev}=\diag{\onev_I\otimes\kav}$, $\tv=\onev_I\otimes[t_1,\ldots,t_T]^{\textsf{T}}$, 
$\X=[\X_1^{\textsf{T}},\ldots,\X_I^{\textsf{T}}]^{\textsf{T}}$, with $\X_i= \onev_T \otimes \xv_i^{\textsf{T}}$ for $i=1,\ldots,I$, $\Z=[\Z_1^{\textsf{T}},\ldots,\Z_I^{\textsf{T}}]^{\textsf{T}}$, with $\Z_i= \onev_T \otimes \zv_i^{\textsf{T}}$ for $i=1,\ldots,I$.
\item Sample 
$$
\gav\mid\text{rest} \sim \textsf{N}_L\le( \V_{\gav}\le[\Z^{\textsf{T}}\SIG^{-1}_{\omev}(\wv-\X\bev-\tv\xi-\tev-\phiv)\ri], \V_{\gav} \ri)
$$
where $\V_{\gav}=(\si_0^{-2}\W^*+\Z^{\textsf{T}}\SIG^{-1}_{\omev}\Z)^{-1}$ and $\W^*$ is a precision matrix of size  $L\times L$ defined accordingly (see Figure \ref{fig_adj_mat}).
\item Sample
$$
\xi\mid\text{rest} \sim \textsf{N}\le( V_{\xi}\le[\tv^{\textsf{T}}\SIG^{-1}_{\omev}(\wv-\X\bev-\Z\gav-\tev-\phiv)\ri], V_{\xi} \ri)
$$
where $V_{\xi}=(\si_0^{-2}+\tv^{\textsf{T}}\SIG^{-1}_{\omev}\tv)^{-1}$.
\item Sample 
$$
\tev_{t}\mid\text{rest} \sim \textsf{N}_I\le( \V_{t}\le[ \ka_t(\omev_{t} - \X_{t}\bev - \Z_{t}\gav - \tv_t\xi  - \phiv_t ) \ri], \V_{t} \ri)
$$
where $\tev_t =(\te_{1t},\ldots,\te_{It})$, $\V_{t}=(\tau_t+\ka_t)^{-1}\I_I$, $\omev_t=(\ome_{1t},\ldots,\ome_{It})$, $\X_t=[\xv_1,\ldots,\xv_I]^{\textsf{T}}$, $\Z_t=[\zv_1,\ldots,\zv_I]^{\textsf{T}} $, and $\tv_t= t\,\onev_I$, for $t=1,\ldots,T$.
\item Sample 
$$
\phiv_{t}\mid\text{rest} \sim \textsf{N}_I\le( \V_{t}\le[ \ka_t(\omev_{t} - \X_{t}\bev - \Z_{t}\gav - \tv_t\xi  - \tev_t ) \ri], \V_{t} \ri)
$$
where $\phiv_t = (\phi_{1t},\ldots,\phi_{It})$ and $\V_{t}=(\lam_t\W +\ka_t\I_I)^{-1}$, for $t=1,\ldots,T$.   
\item Center $\phiv_t$ around its own mean in order to impose the constrain $\sum_{i=1}^I \phi_{i,t}=0$, for $t=1,\ldots,T$.
\item Sample
$
\ka_{t}\mid\text{rest} \sim \textsf{G}\le(({\nu_0+I})/2, \le({\nu_0 + \|\wv_t-\X_t\bev-\Z_t\gav-\tv_t\xi-\tev_t-\phiv_t\|^2}\ri)/2 \ri)
$,
for $t=1,\ldots,T$.
\item Sample
$
\tau_t\mid\text{rest} \sim \textsf{G}\le(a_{\tau}+\tfrac{I}2, b_{\tau} + \tfrac12{\|\tev_t\|^2} \ri)
$,
for $t=1,\ldots,T$.
\item Sample
$
\lam_t\mid\text{rest}\sim \textsf{G}\le(a_{\lam} + \tfrac12{\rank(\W)}, b_{\lam} + \tfrac12{\phiv_t^{\textsf{T}}\W\phiv_t} \ri)
$,
for $t=1,\ldots,T$.
\end{enumerate}

\begin{figure}[ht]
    \centering
    \includegraphics[scale=.5]{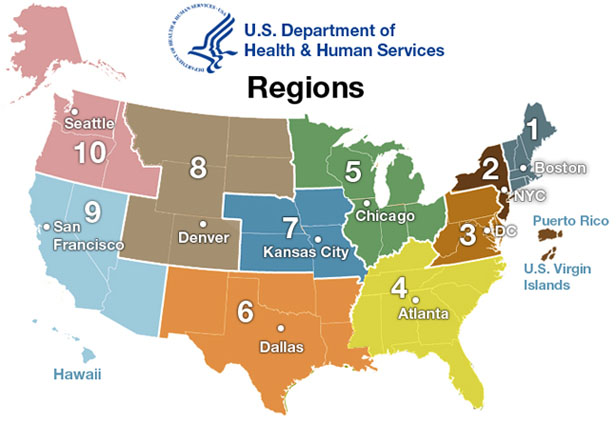}
    \caption{\footnotesize{Regional offices according to the U.S. Department of Health and Human Services..}}\label{fig_map_region}
\end{figure}

Now, we implement the model by considering the Google flu data corresponding to the first 12 weeks of 2013. Outbreaks of influenza are commonly observed during weeks 1 to 8 (Winter season). The proportion of states with Google flue index greater than 7500 cases goes down straight to zero after week 8. Then, we have an align dataset with $I=49$ states (subjects) and $T=12$ weeks (period times) that lead to $637$ observations possibly correlated in time and space. Thus, we consider the linear predictor given by
\begin{equation}\label{eq_predictor}
    F_{\nu_0}^{-1}(\pi_{i,t})= \be_1 + \be_2 x_{i,1} + \be_3 x_{i,2} + \ga_l z_{i,l} + \xi t + \te_{i,t} + \phi_{i,t}
\end{equation}
where $x_{i,1}$ and $x_{i,2}$ are the proportion of white population and the proportion of population over 65 years old, respectively, for state $i$ in 2013, according to Kaiser Family Foundation (\url{http://kff.org/}), and $z_{i,\ell}$ is an indicator variable pointing out if state $i$ belongs to region $\ell$ or not, according to according to the U.S. Department of Health and Human Services (HHS, \url{http://www.hhs.gov/iea/regional/}; see Figure \ref{fig_map_region}). Therefore, $\be_1$ is the mean global effect, $\be_k$ is the fixed effect of covariate $k$, for $k=2,3$, $\ga_\ell$ is the fixed spacial effect of region $\ell$, for $\ell=1,\ldots,10$, $\xi$ is the fixed effect of time, and $\phi_{i,t}+\phi_{i,t}$ is the random spatiotemporal effect considering heterogeneity and clustering.

We implement the MCMC algorithm given above drawing $B = 25000$ samples from the posterior distribution $p(\mathbf{\Upsilon}\mid\boldsymbol{y})$ after thinning the original chain every 10 observations and a burn-in period of $5000$ iterations. To do so, we weakly concentrate the prior distribution of $\be_k$ , $\ga_\ell$, and $\xi$ around 0, and that of $\ka_t$, $\tau_t$, and $\lam_t$ around 1, by following the prior specification given in (\ref{eq_prior_elicitation}) with $\si^2_0 = 100$, $\nu_0=a_{\tau}=b_{\tau}=a_{\lam} =2$, and $b_{\lam}=(0.7^2)\bar{m}b_\tau$. Notice that $\nu_0$ is set equal to 2 because a $\textsf{t}_2$ distribution has an infinite variance. Trace and autocorrelation plots of the regression parameters show that the corresponding chains achieve convergence quickly. However, in some particular cases, there are signs of serious autocorrelation (in order to address this issue we thin the chain as specified).

\begin{figure}[t]
    \centering
    \includegraphics[scale=.62]{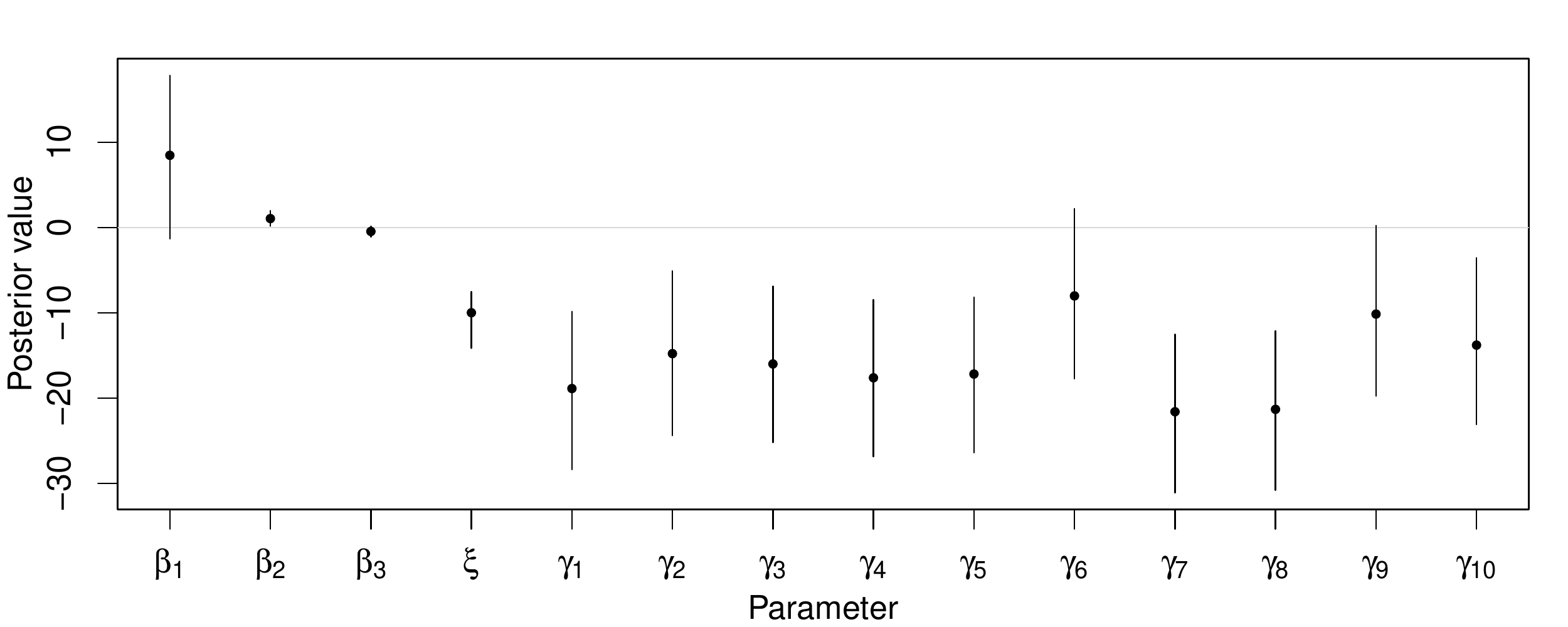}
    \caption{\footnotesize{Posterior mean and 95\% quantile-based credible intervals for $\bev$, $\xi$, and $\gav$.}}\label{fig_post_bg}
\end{figure}

Figure \ref{fig_post_bg} summarizes the posterior distribution of $\bev$, $\xi$, and $\gav$. The proportion of the white population have a positive significant impact on the response, which leads to an increment in the probability that the state is above the threshold, whereas the global mean and the proportion of population over 65 do not. Even thought most deaths associated with influenza in industrialized countries occur among the elderly over 65 years of age, this covariate results not significant because it has little variation across the states ($\textsf{CV}\doteq11\%$) and the proportion of population over 65 in each state is just a small fraction of the whole state population (be aware of the ``ecological fallacy''). Furthermore, all the regions delimited by the HHS have a significant impact on the response, excepting regions 6 and 9 (see Figure \ref{fig_map_region}). This finding is reasonable since these regions are neighbors and are the smallest regions in terms of extension and population. In addition, it is not surprising that in general regions have a negative influence on the response since these regions independently address needs of communities and individuals through HHS programs and policies. Finally, the effect of time is also significant since it is a well known that outbreaks of flue are highly correlated with seasons and we consider a time transition frame.

\begin{figure}[ht]
    \centering
    \includegraphics[scale=.62]{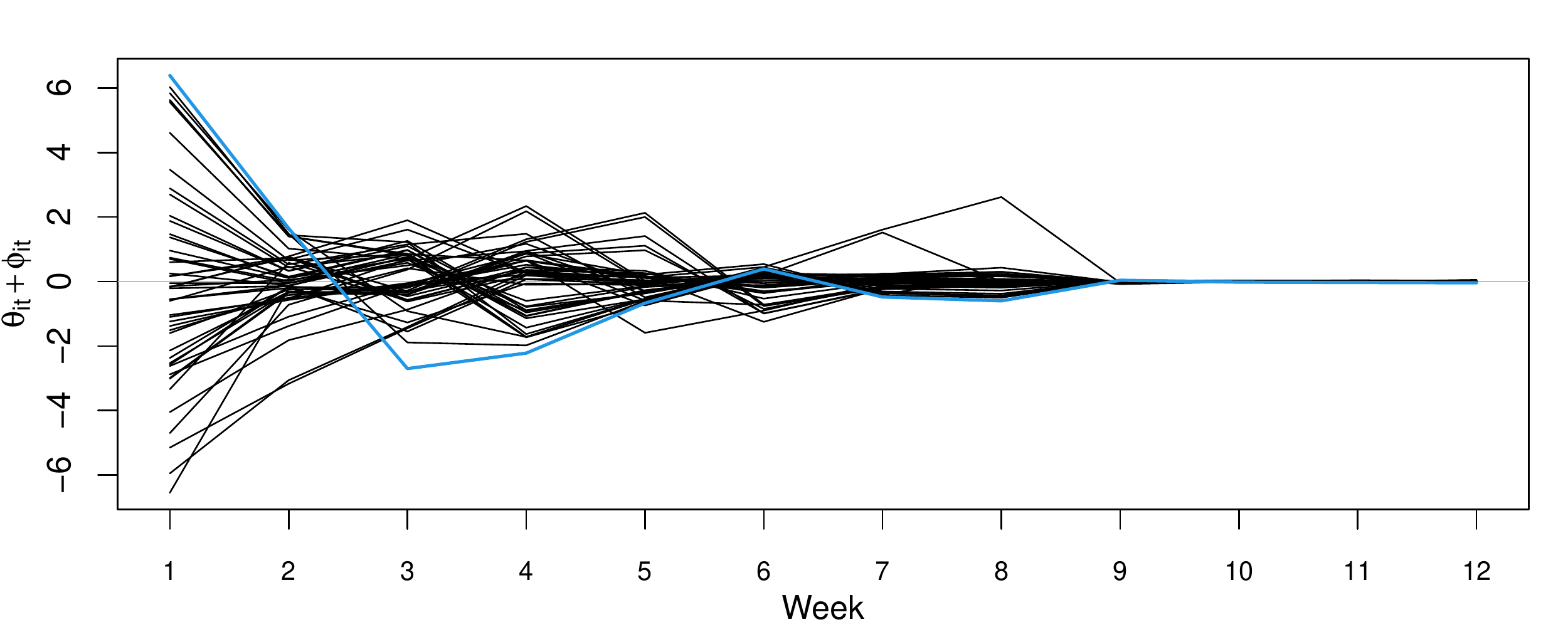}
    \caption{\footnotesize{Posterior mean for the dynamic random effects $\te_{i,t}+\phi_{i,t}$. The blue line corresponds to Maine, which is the state with the highest proportion of white population in 2013.}}\label{fig_post_ran_effects}
\end{figure}

On the other hand, Figure \ref{fig_post_ran_effects} displays the posterior mean for the sum of the heterogeneity and clustering random effects $\te_{i,t}+\phi_{i,t}$. As expected, these effects clearly go down toward 0 after week 8 due to the seasonal behavior of the response. Even though all the states reveal similar patterns, the model is also useful to identify subject-specific dynamic effects. For instance, consider the case of Maine (ME), which is the state with the highest proportion of white population in 2013.

%%%%%%%%%%%%%%%%%%%%%%%%%%%%%%%%%%%%%%%%%%%%%%%%%%%%%%%%%%%%%%%%%%%%%%%%%%%%%%%%%%%%%%%%%%%%%%%%%%%%%%%%%%%%%%%%%%%%
\section{Species sampling modeling}\label{sec_species}

We consider the species sampling model (SSS)
$$
p(X_{n+1}\in B\mid X_1,\ldots,X_n) = \sum_{k=1}^{K^n} q^n_k(\mv^n)\del_{\tilde{X}_k}(B) + q^n_{K^n+1}(\mv^n)G_0(B)
$$
where $G_0$ is a non-atomic measure \citep{rodriguez2015species}. First of all, in order to understand and develop the MCMC algorithms presented below, we find an expression for $\expec{K^n}$, the expected value of the number of distinct species among $X_1,\ldots,X_n$.

Let $W_1,\ldots,W_n$ be a sequence of binary random variables such that
$$
W_i:=\left\{
       \begin{array}{ll}
         1, & \hbox{$X_i$ corresponds to a species that has not been observed in $X_1,\ldots,X_{i-1}$;}\\
         0, & \hbox{otherwise,}
       \end{array}
     \right.
$$
for $i=1,\ldots,n$. Since the first observation is necessarily a new observation we have that $W_1\equiv1$ and
$$
\expec{K^n}=\expec{\sum_{i=1}^n W_i} = \sum_{i=1}^n \expec{ W_i} = 1 + \sum_{i=2}^n \expec{ W_i}.
$$

Let $\mathcal{M}_k$ denote the set of all vectors of length $k$ with positive, integer entries that add up to $n$, i.e., $\mathcal{M}_k:=\{(m_1,\ldots,m_k)\in \mathbb{N}^k: m_1+\ldots+m_k=n\}$. Case by case, we see that
\begin{align*}
\expec{W_2}&=\sum_{\mathcal{M}_1} p(m^1_1,1)\\
\expec{W_3}&=\sum_{\mathcal{M}_1} p(m^2_1,1) + \sum_{\mathcal{M}_2} p(m^2_1,m^2_2,1)  \\
\expec{W_4}&=\sum_{\mathcal{M}_1} p(m^3_1,1) + \sum_{\mathcal{M}_2} p(m^3_1,m^3_2,1) + \sum_{\mathcal{M}_3} p(m^3_1,m^3_2,m^3_3,1)  \\
&\hspace{.15cm}\vdots\\
\expec{W_i}&=\sum_{\mathcal{M}_1} p(m^{i-1}_1,1) + \sum_{\mathcal{M}_2} p(m^{i-1}_1,m^{i-1}_2,1) + \ldots + \sum_{\mathcal{M}_{i-1}} p(m^{i-1}_1,\ldots,m^{i-1}_{i-1},1)  \\
\end{align*}
and therefore
$$
\expec{W_i}=\sum_{k=1}^{i-1}\sum_{\mathcal{M}_k} p(m^{i-1}_1,\ldots,m^{i-1}_{k},1).
$$
Recalling the following fact given in \citet[p. 152]{pitman1995exchangeable},
$$
p(m^{i-1}_1,\ldots,m^{i-1}_{k},1) = p(m^{i-1}_1,\ldots,m^{i-1}_{k})\,q_{k+1}^{i-1}(m^{i-1}_1,\ldots,m^{i-1}_{k})\,,\quad i=2,\ldots,n,
$$
we finally get that the expected value of the number of distinct species is given by
$$
\expec{K^n} = 1 + \sum_{i=2}^n\sum_{k=1}^{i-1}\sum_{\mathcal{M}_k}p(m^{i-1}_1,\ldots,m^{i-1}_{k})\,q_{k+1}^{i-1}(m^{i-1}_1,\ldots,m^{i-1}_{k})\,.
$$

In order to verify this result, we consider two particular scenarios, namely, the Chinese restaurant process (CRP) and the Pitman-Yor process (PYP):
\begin{description}
\item[CRP]
In this case, the predictive probability function (PPF) is given by
$q_{k+1}^{i-1}(m_1^{i-1},\ldots,m_k^{i-1}) = \frac{\te}{\te+i-1}$
\cite[p. 83]{muller2013nonparametric}, and therefore
$$
\expec{K^n} = 1 + \sum_{i=2}^n\frac{\te}{\te+i-1} \sum_{k=1}^{i-1}\sum_{\mathcal{M}_k} p(m^{i-1}_1,\ldots,m^{i-1}_{k})
= 1 + \sum_{i=2}^n\frac{\te}{\te+i-1}
$$
since
$$
\sum_{k=1}^{i-1}\sum_{\mathcal{M}_k} p(m^{i-1}_1,\ldots,m^{i-1}_{k}) = 1\,,\quad i=1,\ldots,n,
$$
which leads to
$$
\expec{K^n} =\sum_{i=1}^n\frac{\te}{\te+i-1}
$$
This expression is the same one given in \citet[p. 6]{rodriguez2015species}.

\item[PYP]
In this case the PPF is given by $q_{k+1}^{i-1}(m_1^{i-1},\ldots,m_k^{i-1}) = \frac{\te + \sig k}{\te+i-1}$ \cite[p. 85]{muller2013nonparametric}, which means that
\begin{align*}
\expec{K^n} &= 1 + \sum_{i=2}^n\sum_{k=1}^{i-1}\sum_{\mathcal{M}_k} p(m^{i-1}_1,\ldots,m^{i-1}_{k})\,\frac{\te+\si k}{\te+i-1}\\
&= 1 + \sum_{i=2}^n\frac{\te}{\te+i-1}\sum_{k=1}^{i-1}\sum_{\mathcal{M}_k} p(m^{i-1}_1,\ldots,m^{i-1}_{k}) \\
&\hspace{2cm}+\sum_{i=2}^n\frac{\si}{\te+i-1}\sum_{k=1}^{i-1}\sum_{\mathcal{M}_k} k\, p(m^{i-1}_1,\ldots,m^{i-1}_{k})\\
\end{align*}
and therefore
$$
\expec{K^n} = \sum_{i=1}^n\frac{\te}{\te+i-1} +\sum_{i=2}^n\frac{\si}{\te+i-1}\,\zeta_i
$$
with
$$\zeta_i:=\sum_{k=1}^{i-1}\sum_{\mathcal{M}_k} k\, p(m^{i-1}_1,\ldots,m^{i-1}_{k})\,,\quad i=2,\ldots,n.$$
\end{description}

Now, consider the stick breaking prior of the form $G=\sum_{l=1}^\infty \ome_l\del_{X^*_l}$ where $X_l^*\simiid G_0$, $\om_l=z_l\prod_{k<l}(1-z_k)$, and $z_l\simiid H^{\tev}$ for all $l=1,2\ldots$, with $H^{\tev}$ a probability distribution on $[0,1]$ indexed by the vector of parameters $\tev$. In what follows we show that the exchangeable partition probability function (EPPF) induced by this stick breaking prior is given by
\begin{equation}\label{eq_theorem_prob_3}
p(m_1^n,\ldots,m_{K^n}^n \mid \tev)=\sum_{\siv\in\mathcal{P}_{K^n}}\le[ \prod_{k=1}^{K^n} \frac{\ga_{\tev}\le(m^n_{\si_k},\sum_{j=k+1}^{K^n}m^n_{\si_j}\ri)}{1-\ga_{\tev}\le(0,\sum_{j=k}^{K^n}m^n_{\si_j}\ri)}  \ri]
\end{equation}
where $\mathcal{P}_{K^n}$ denotes the set of all permutations of $\{1,\ldots,K^n\}$, $\siv=(\si_1,\ldots,\si_{K^n})$, and $\ga_{\tev}(x,y)=\expec{z^x(1-z)^y}$ with $z\sim H^{\tev}$.

If $X_1,\ldots,X_n$ is a iid sample from $G$, then this sequence is exchangeable and there exists a positive probability of ties among the $X_i$\,s. Furthermore, the EPPF associated with this prior is given by
\begin{equation}\label{eq_EPPF_prob_3}
p(\mv^n \mid \tev) = \sum_{(j_1,\ldots,j_{K^n})\in \mathfrak{I}_{K^n}} \expec{\prod_{k=1}^{K^n}\om_{j_k}^{m^n_k}}
\end{equation}
where $\mv=(m_1^n,\ldots,m_{K^n}^n)$ and $I_{K^n}$ is the set of all possible sequences of distinct positive integers of length $K^n$ \cite[p. 256]{pitman1996some}. Then, by expanding the summation, the previous expression becomes
$$
p(\mv^n \mid \tev) = \sum_{j_1}\sum_{j_2\notin\{j_1\}}\sum_{j_3\notin\{j_1,j_2\}}\ldots\sum_{j_{K^n}\notin\{j_1,\ldots,j_{K^n-1}\}} \expec{ \prod_{k=1}^{K^n} \le( z_{j_k}\prod_{h<j_k} (1-z_h) \ri)^{m^n_k} }\,.
$$

First, we consider the scenario in which $j_1<j_2<\ldots<j_{K^n}$. Therefore,
{\footnotesize
\begin{align*}
\prod_{k=1}^{K^n} \le( z_{j_k}\prod_{h<j_k} (1-z_h) \ri)^{m^n_k}
&= \prod_{k=1}^{K^n} z_{j_k}^{m^n_k} (1-z_1)^{m^n_k}(1-z_2)^{m^n_k}\ldots(1-z_{j_k-1})^{m^n_k} \\
&=\le[z_{j_1}\ldots z_{K^n}\ri]^{S_1^n}\cdot \le[(1-z_1)\ldots(1-z_{j_1-1})\ri]^{S_1^n}\cdot\le[(1-z_{j_1})\ldots(1-z_{j_2-1})\ri]^{S_2^n}\\
&\quad\,\cdot\le[(1-z_{j_2})\ldots(1-z_{j_3-1})\ri]^{S_3^n}\cdot\ldots\cdot\le[(1-z_{j_{(K^n-1)}})\ldots(1-z_{j_{(K^n)}-1})\ri]^{S_{K^n}^n}
\end{align*}
}
where $S_k^n=\sum_{j=k}^{K^n} m^n_j$ for $k=1,\ldots,K^n$.

Associating terms and recalling the fact that $z_l\iid H^{\tev}$ for $l=1,2,\ldots$ we get that
$$
 \expec{\prod_{k=1}^{K^n}\om_{j_k}^{m^n_k}}= \prod_{k=1}^{K^n} \expec{z^{m_k}(1-z)^{S^n_{k+1}}} \le(\expec{(1-z)^{S_k}}\ri)^{d_k}
$$
where $z\sim H^{\tev}$ and $d_k=j_k-j_{k-1}-1$ for $k=1,\ldots,K^n$, which represent the sizes of the ``gaps'' between he labels of two consecutive observed species, with the convention $j_0=0$ \cite[p. 14]{rodriguez2015species}.

Now, recognizing the symmetry of the summation in (\ref{eq_EPPF_prob_3}) and all the possible ``gaps'' we obtain that
\begin{align*}
p(\mv^n \mid \tev) &= \sum_{d_1=0}^\infty\ldots\sum_{d_{K^n}=0}^\infty \expec{\prod_{k=1}^{K^n}\om_{j_k}^{m^n_k}}\\
&=\prod_{k=1}^{K^n} \le\{\expec{z^{m_k}(1-z)^{S^n_{k+1}}} \sum_{d_k=0}^\infty\le(\expec{(1-z)^{S_k}}\ri)^{d_k} \ri\}
\end{align*}
Since the las summation is a geometric series then we finally get that
$$
p(\mv^n \mid \tev) =
\prod_{k=1}^{K^n} \frac{ \expec{z^{m_k}(1-z)^{S^n_{k+1}}}}{1- \expec{(1-z)^{S_k}}}
$$
which precisely corresponds to the expression given in  (\ref{eq_EPPF_prob_3}).

We consider the data discussed in \citet[Sec. 4.2]{lijoi2007bayesian}. The basic sample consists of $n = 2586$ expressed sequence tags and this gives $K^n=1825$ different CDNA fragments each of which represents a unique gene. If $r_i$ denotes the number of clusters of size $i$, then the dataset gives $r_i = 1434,253,71,33,11,6,2,3,1,2,2,1,1,1,2,1,1$ with $i \in \{1, 2,..., 14\} \cup \{16, 23, 27)\}$. This means we are observing 1434 clusters of size 1, 253 clusters of size 2, and so on.

Here, we develop MCMC algorithms for performing Bayesian estimation of the parameters of a CRP and a PYP using this dataset. These algorithms were design according to the posterior distribution
$$
p(\tev|\mv^n) \propto p(\mv^n|\tev)\,p(\tev)
$$
where $\tev$ is the vector of parameters of the process and $\mv^n=(m_1^n,\ldots,m_{K^n}^n)$. Below we describe the algorithms for performing Bayesian estimation of $\tev$. Let $\tev^{(m)}$ denote the parameter vector at iteration $m$ of the algorithm, $m=0,1,\ldots,M$. Given a starting point of the parameter $\tev^{(0)}$, we run the sampler updating $\tev^{(m-1)}$ to $\tev^{(m)}$ until convergence according to the corresponding description.

\begin{description}
  \item[CRP] Here $\tev=\te$ and
$$
p(\te|\mv^n) \propto \le[\te^{K^n}\frac{\Ga(\te)}{\Ga(\te+n)}\,\prod_{j=1}^{K^n}\Ga(m_j^n)\ri]\times \textsf{G}(\te|a_\te, b_\te)\,.
$$
The MCMC algorithm is as follows:
    \begin{enumerate}[1.]
    \item Compute $\eta^{(m-1)}=\log\te^{(m-1)}$.
    \item Sample $\eta^*\sim \textsf{N}(\eta^{(m-1)},b_\eta)$, where $b_\eta$ is a tuning parameter.
    \item Compute $r=p_\eta(\eta^*|\mv^n)/p_\eta(\eta^{(m-1)}|\mv^n)$ where $p_\eta(\eta|\mv^n) = p(e^{\eta}|\mv^n)\times e^{\eta}$.
    \item Set
        $$\eta^{(m)}=\left\{
                       \begin{array}{ll}
                         \eta^*, & \hbox{with probability $r$;} \\
                         \eta^{(m-1)}, & \hbox{with probability $1-r$.}
                       \end{array}
                     \right.
        $$
   \item Compute $\te^{(m)}=\exp{ \left\{ \eta^{(m)} \right\} }$.
    \item Repeat until convergence.
    \end{enumerate}

Notice that in the Metropolis step above we do the transformation $\eta=\log(\te)$ to consider the range of $\te$ in the random-walk proposal. The corresponding Jacobian ($e^\eta$) is taken into account in the computations through $p_\eta(\eta|\mv^n)$.

   \begin{figure}[h!]
        \centering
        \subfigure[Mixing rate]{\includegraphics[scale=.5]{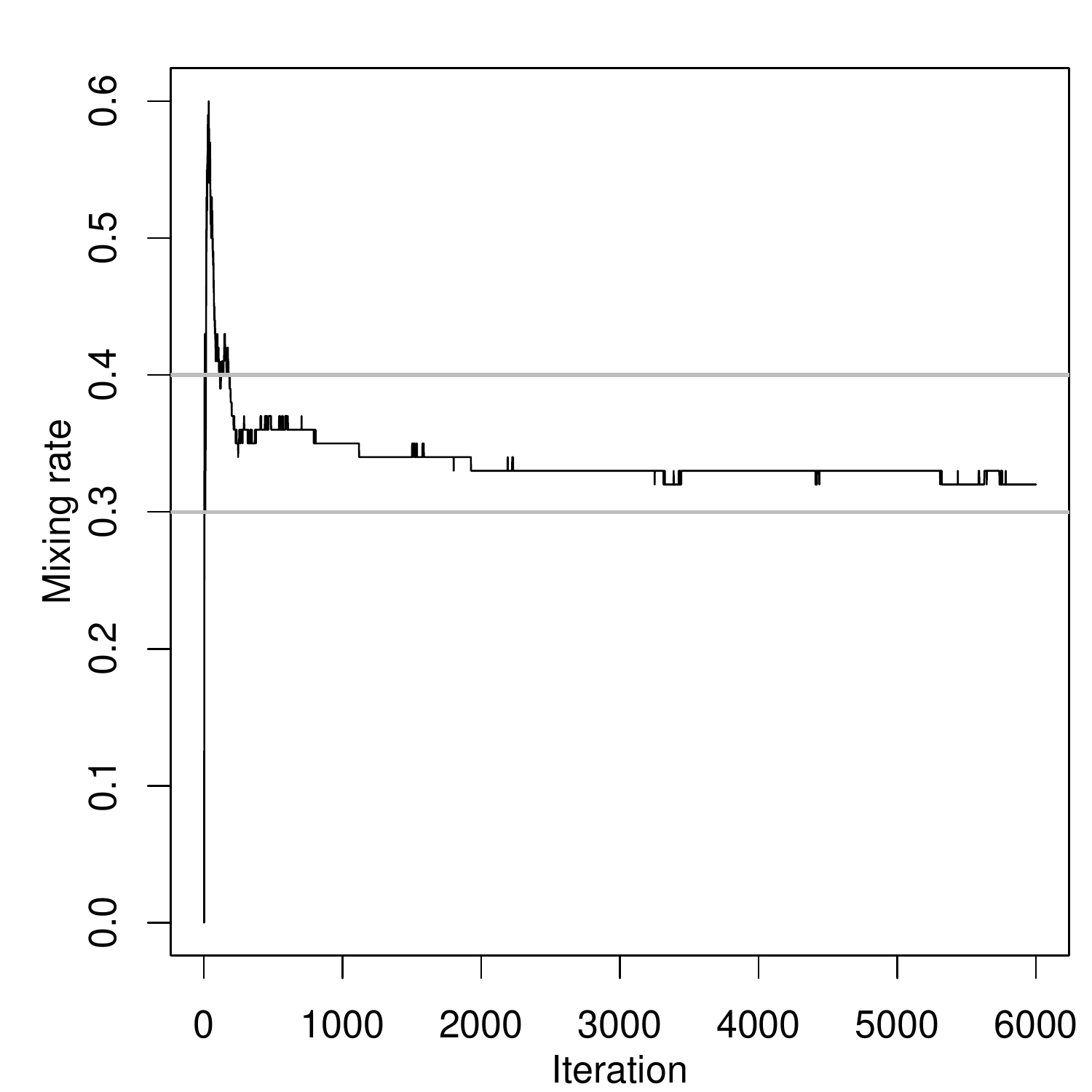}}
        \subfigure[Traceplot]  {\includegraphics[scale=.5]{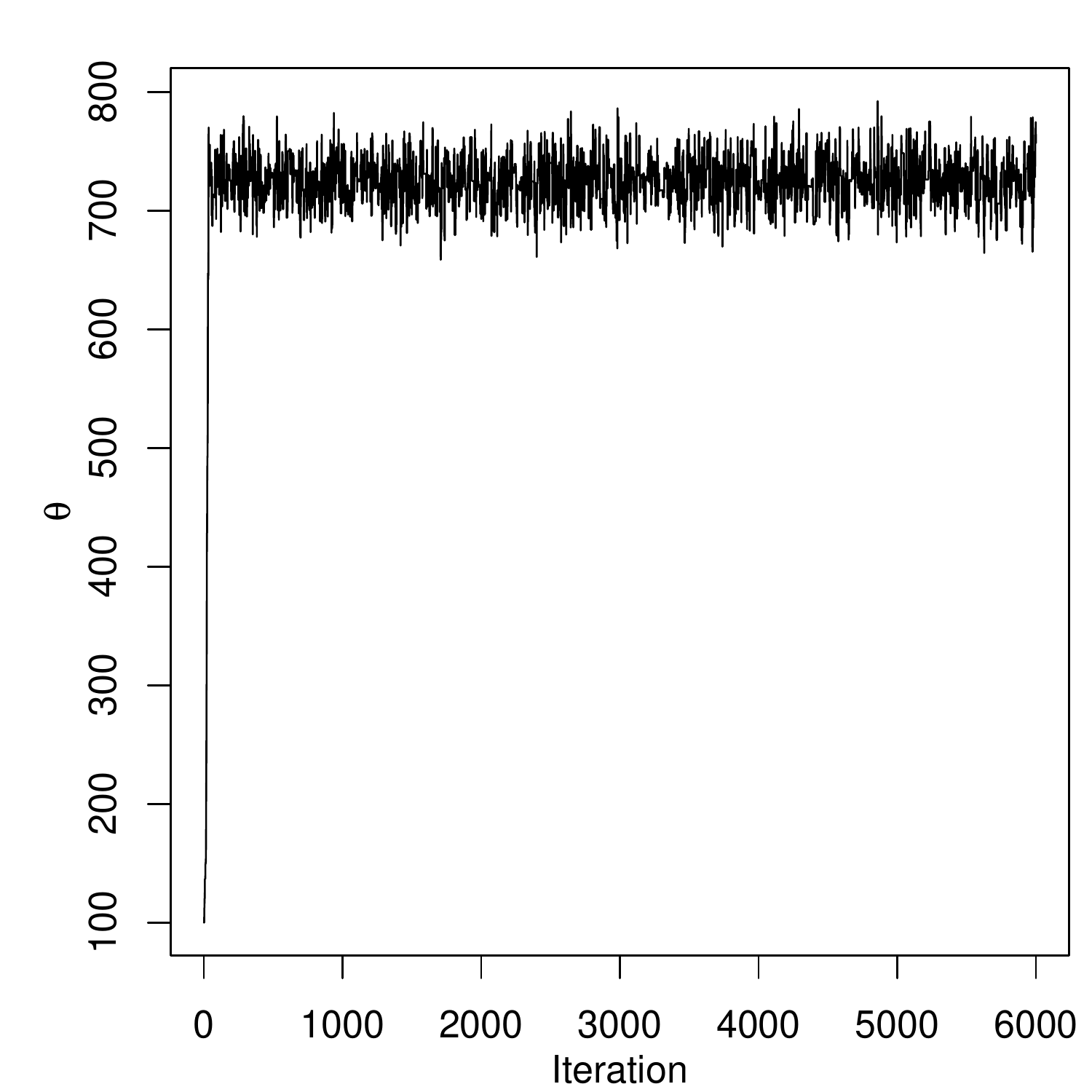}}
        \caption{Mixing rates associated with the Metropolis algorithm and traceplot for $\te$ in the CRP.}\label{fig_CRP_CODA}
   \end{figure}

We run the sampler with $M=6000$ iterations using as hyperparameters $a_\te=1$ and $b_\te=1$, and $b=0.1$ as tuning parameter (which leads to nice mixing rates between 30\% and 40\%, see left panel of Figure \ref{fig_CRP_CODA}). In addition, we use a burn-in period with 1000 iterations and, in order to have approximately independent draws, select one sample from each 50 iterations. As a consequence a total of 100 samples are selected altogether (the chain quickly achieves convergence, see right panel of Figure \ref{fig_CRP_CODA}). Those 100 samples are approximately independent and identically distributed according to the corresponding posterior distribution and form the basis of posterior inference. Finally, table \ref{tab_CRP_POST_theta} summarizes the posterior distribution of $\te$.

\begin{table}[ht]
\centering
\begin{tabular}{lccccc}
  \hline
 Parameter & Mean & Median & SD & Q\,2.5\% & Q\,97.5\% \\
  \hline
$\te$  & 723.42 & 724.04 & 22.13 & 679.90 & 768.25 \\
   \hline
\end{tabular} \caption{\footnotesize{Posterior summaries for $\te$ in the CRP.}}\label{tab_CRP_POST_theta}
\end{table}

\item[PYP] Here $\tev=(\te,\si)$ and
\begin{align*}
p(\te|\mv^n) &\propto \le[\frac{\Ga(\te+1)}{\Ga(\te+K^n\si)\Ga(\te+n)}\,\prod_{j=1}^{K^n}(\te+j\si)\frac{\Ga(m_j^n-\si)}{\Ga(1-\si)}\ri]\\
&\hspace{0.5cm}\times \textsf{NT}_{(-\si,\infty)}(\te|a_\te,b_\te) \times \textsf{Be}(\si|a_\si, b_\si)
\end{align*}
The MCMC algorithm is as follows:
    \begin{enumerate}[1.]
    \item Update $\si^{(m-1)}$ given $\te^{(m-1)}$ and $\mv^n$ as follows:
    \begin{enumerate}[i.]
    \item Compute $\eta^{(m-1)}=\logit\si^{(m-1)}$, with $\logit x=\log\frac{x}{1-x}$.
    \item Sample $\eta^*\sim N(\eta^{(m-1)},b_\eta)$, where $b_\eta$ is a fixed positive constant.
    \item Compute $r=p_\eta(\eta^*|\te^{(m-1)},\mv^n)/p_\eta(\eta^{(m-1)}|\te^{(m-1)},\mv^n)$ where
            $$p_\eta(\eta|\te,\mv^n) = p(\logitinv\eta|\te,\mv^n) \times \frac{e^\eta}{(e^\eta+1)^2}$$
          with $\logitinv x=\frac{e^x}{e^x+1}$.

    \item Set
        $$\eta^{(m)}=\left\{
                       \begin{array}{ll}
                         \eta^*, & \hbox{with probability $r$;} \\
                         \eta^{(m-1)}, & \hbox{with probability $1-r$.}
                       \end{array}
                     \right.
        $$
   \item Compute $\si^{(m)}= \logitinv\eta^{(m)}$.
    \end{enumerate}
%///
    \item Update $\te^{(m-1)}$ given $\si^{(m)}$ and $\mv^n$ as follows:
    \begin{enumerate}[i.]
    \item Compute $\psi^{(m-1)}=\log(\te^{(m-1)}+\si^{(m)})$.
    \item Sample $\psi^*\sim N(\psi^{(m-1)},b_\psi)$, where $b_\psi$ is a fixed positive constant.
    \item Compute $s=p_\psi(\psi^*|\si^{(m)},\mv^n)/p_\psi(\psi^{(m-1)}|\si^{(m)},\mv^n)$, where
            $$p_\psi(\psi|\te,\mv^n) = p(e^\psi-\si|\te,\mv^n) \times e^\psi\,.$$

    \item Set
        $$\psi^{(m)}=\left\{
                       \begin{array}{ll}
                         \psi^*, & \hbox{with probability $s$;} \\
                         \psi^{(m-1)}, & \hbox{with probability $1-s$.}
                       \end{array}
                     \right.
        $$
   \item Compute $\te^{(m)}= \ex{\psi^{(m)}}-\si^{(m)}$.
    \end{enumerate}
    \item Repeat until convergence.
    \end{enumerate}

Notice that in the Metropolis steps above we do two transformations to consider the range of parameters in the random-walk proposals. The corresponding Jacobians are taken into account in the computations.

\begin{figure}[h!]
        \centering
        \subfigure[Traceplot for $\sigma$]{\includegraphics[scale=.5]{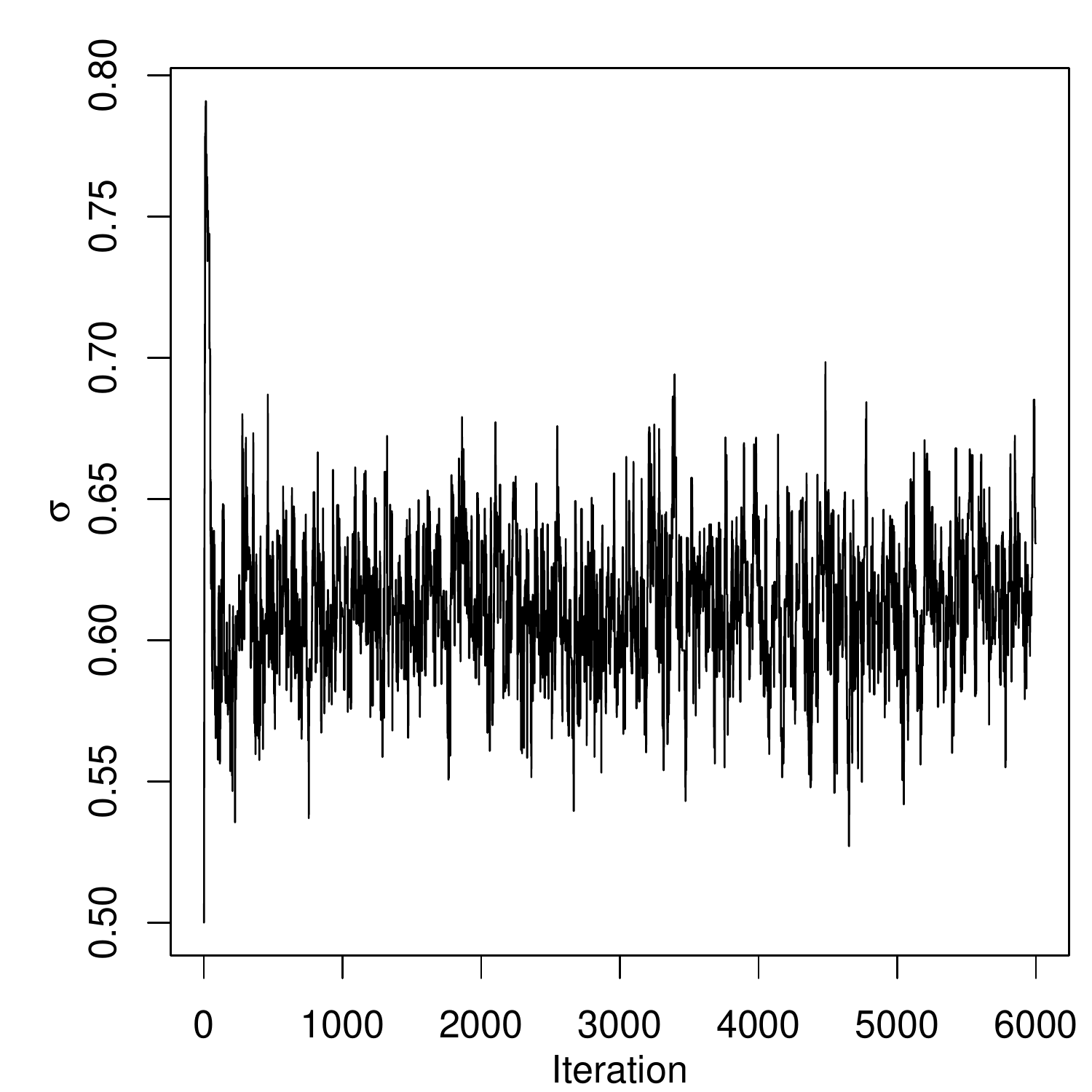}}
        \subfigure[Traceplot for $\theta$]{\includegraphics[scale=.5]{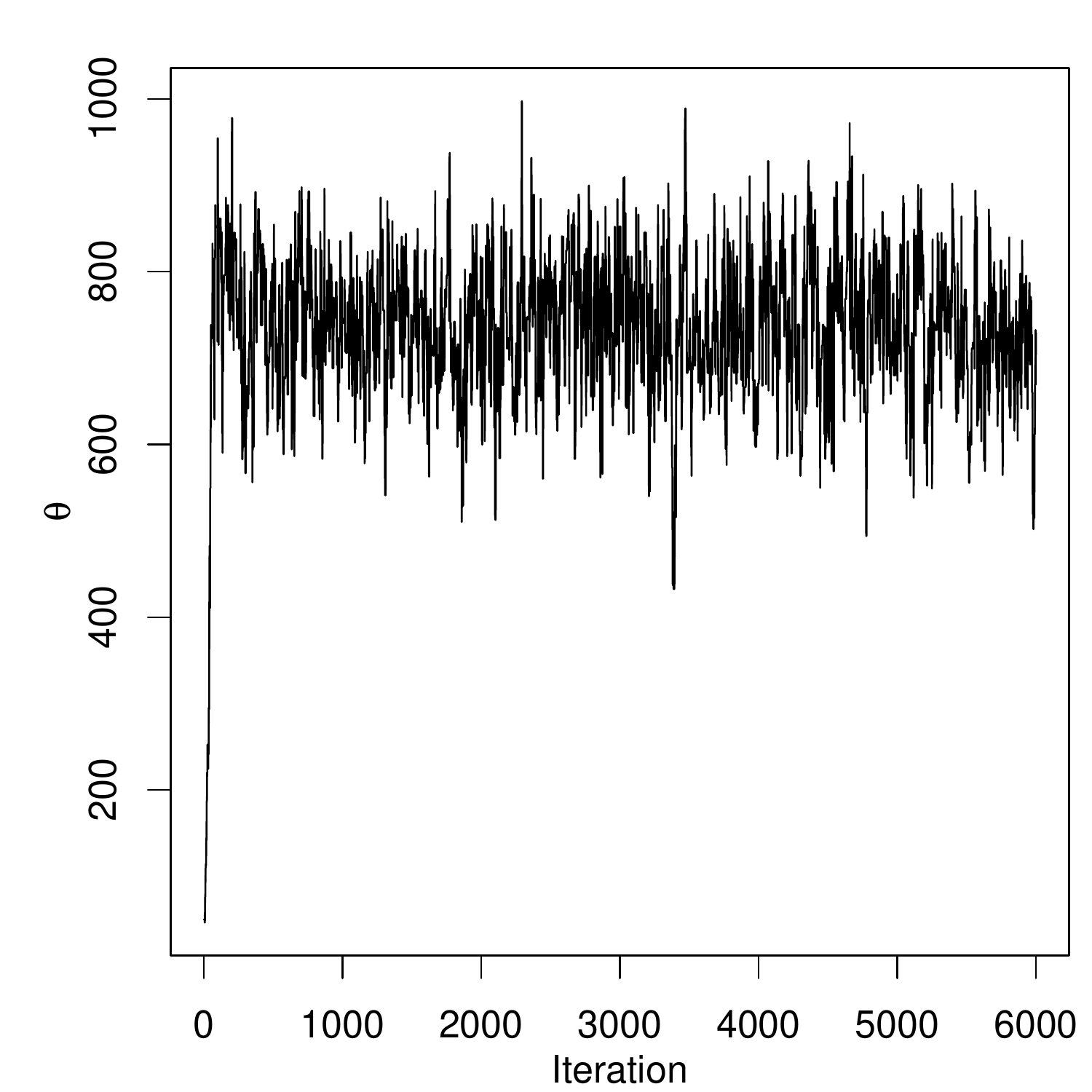}}
        \caption{Traceplots for the $\sigma$ and $\theta$ in the PYP.}\label{fig_PYP_CODA}
   \end{figure}

Once again we run the sampler with $M=6000$ iterations using as hyperparameters $a_\si=1$, $b_\si=1$, $a_\te=723$ and $b_\te=100$ (fairly uninformative prior distributions), and $b_\eta=0.2$ and $b_\psi=.2$ as tuning parameters (which leads to nice mixing rates between 30\% and 40\% not shown here). Again, we use a burn-in period with 1000 iterations and select one sample from each 50 iterations giving as a result a total of 100 samples altogether (the chains quickly achieve convergence, see Figure \ref{fig_PYP_CODA}). Finally, table \ref{tab_PYP_POST_theta} summarizes the posterior distribution of $\si$ and $\te$. This values are certainly closed to the corresponding frequentist estimates ($\hat{\si}_\mle=0.612$ and $\hat{\te}_\mle=741$) \cite[p. 778]{lijoi2007bayesian}.

\begin{table}[ht]
\centering
\begin{tabular}{lccccc}
  \hline
 Par. & Mean & Median & Sd & Q\,2.5\% & Q\,97.5\% \\
  \hline
$\si$  & 0.61 & 0.61 & 0.03 & 0.56 & 0.66 \\
$\te$  &  733.01 & 739.00 & 89.79 & 569.37 & 876.31 \\
   \hline
\end{tabular} \caption{\footnotesize{Posterior summaries for $\si$ and $\te$ in the PYP.}}\label{tab_PYP_POST_theta}
\end{table}
\end{description}

Now we extend the previous algorithms to produce predictions on the number of new species in a new sample of size $n^*$. We consider two extensions: one using purely simulations from the SSM, and other one using the formulas discussed in \citet{lijoi2007bayesian}. First of all, recall that
\begin{equation}\label{eq_EPPFtimesPPF_1}
p(m_1^n,\ldots,m^n_k+1,\ldots,m^n_{K^n}|\tev) = p(\mv^n|\tev)\,q^n_k(\mv^n|\tev)\,,\quad k\leq K^n\,,
\end{equation}
and
\begin{equation}\label{eq_EPPFtimesPPF_2}
p(m_1^n,\ldots,m^n_{K^n},1|\tev) = p(\mv^n|\tev)\,q^n_{K^n+1}(\mv^n|\tev)
\end{equation}
where $p(\mv^n)$ is the EPPF, $q^n_{k}(\mv^n)$, $k=1,\ldots,K^n+1$, is the corresponding PPF, and $\mv^n=(m_1^n,\ldots,m^n_{K^n})$. Thus, we apply the following algorithm based on direct simulation from the SSM in order to produce predictions on the number of new species:
    \begin{enumerate}[1.]
    \item For each $s=1,\ldots,n^*$:
    \begin{enumerate}[i.]
      \item Compute the distribution given by (\ref{eq_EPPFtimesPPF_1}) and (\ref{eq_EPPFtimesPPF_2}) for  $k=0,1,\ldots,K^s+1$.
      \item Sample $k^*$ from $\{0,1,\ldots,K^s+1\}$ according to the previous distribution.
      \item Update
        $$
        \mv^{s+1}=\left\{
                    \begin{array}{ll}
                      \mv^s, & \hbox{if $1\leq k^*\leq K^s$;} \\
                      (\mv^s,1), & \hbox{if $k^*=K^s+1$,}
                    \end{array}
                  \right.
        $$
        and
        $$
        K^{s+1}= \left\{
                \begin{array}{ll}
                  K^s, & \hbox{if $1\leq k^*\leq K^s$;} \\
                  K^s+1, & \hbox{if $k^*=K^s+1$.}
                \end{array}
              \right.
        $$
    \end{enumerate}
    \item Set $K^{(m)}=K^{n^*+1}$.
    \item Repeat for every $\tev^{(m)}$ in the chain for $m=1,\ldots,M$.
    \end{enumerate}

On the other hand, considering the formulas to estimating the probability of discovering a new species given in \citet[Sec. 3]{lijoi2007bayesian}, we implement the following algorithm to produce predictions on the number of new species:
    \begin{enumerate}[1.]
     \item Compute the distribution $\pr{K_{n^*}=k|\te,\mv^n}$ for  $k=0,1,\ldots,n^*$, where $K_{n^*}:=K^{n+n*}-K^n$ is the number of new species in the second (hypothetic) sample of size $n*$, by using to the specific formulas given in \citet[Sec. 3]{lijoi2007bayesian}.
     \item Sample $k^*$ from $\{0,1,\ldots,K^s+1\}$ according to the previous distribution.
     \item Set $K^{(m)}=k^*$.
     \item Repeat for every $\tev^{(m)}$ in the chain for $m=1,\ldots,M$.
    \end{enumerate}

\begin{figure}[h]
        \centering
        \subfigure[SSM based]{\includegraphics[scale=.5]{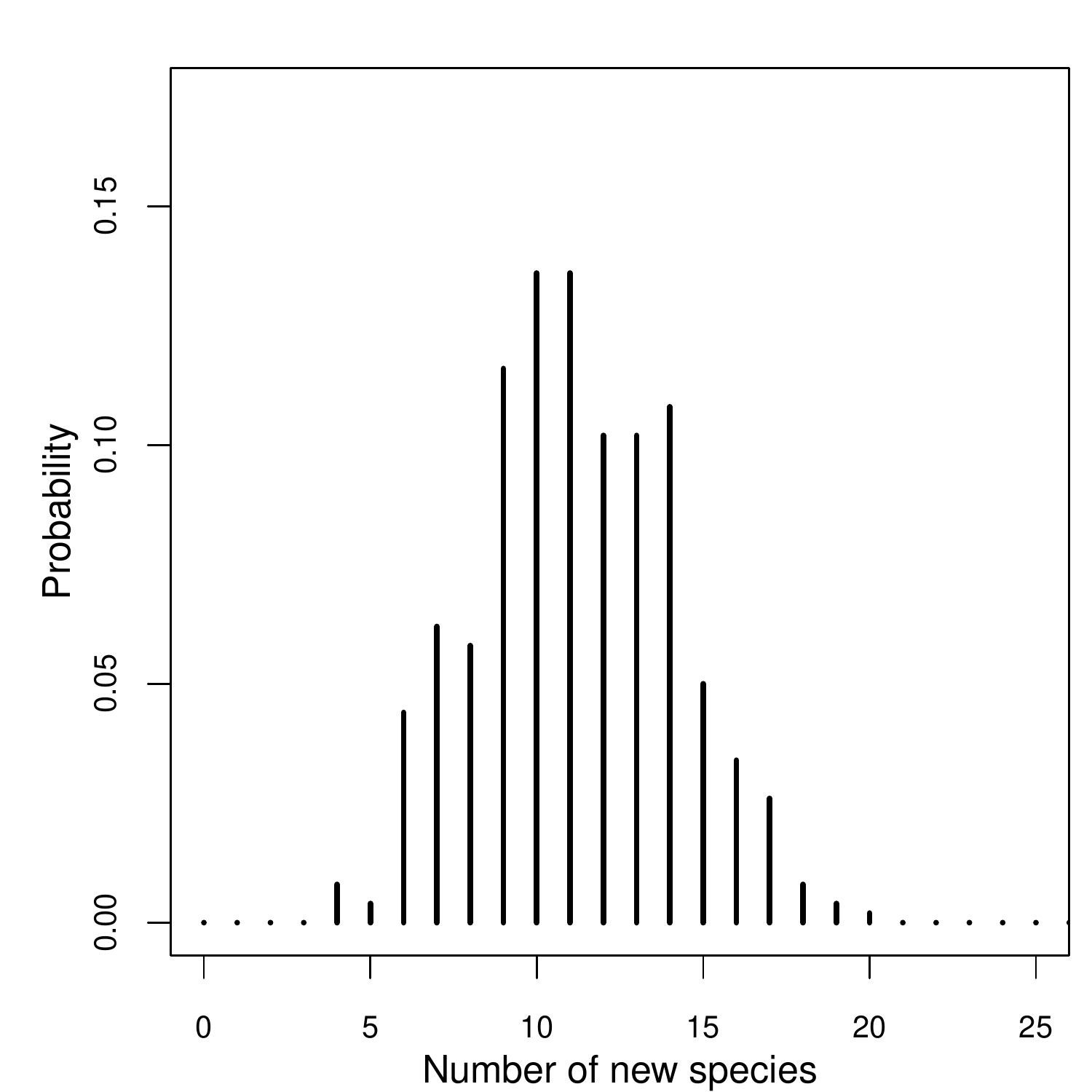}}
        \subfigure[Lijoi’s based]{\includegraphics[scale=.5]{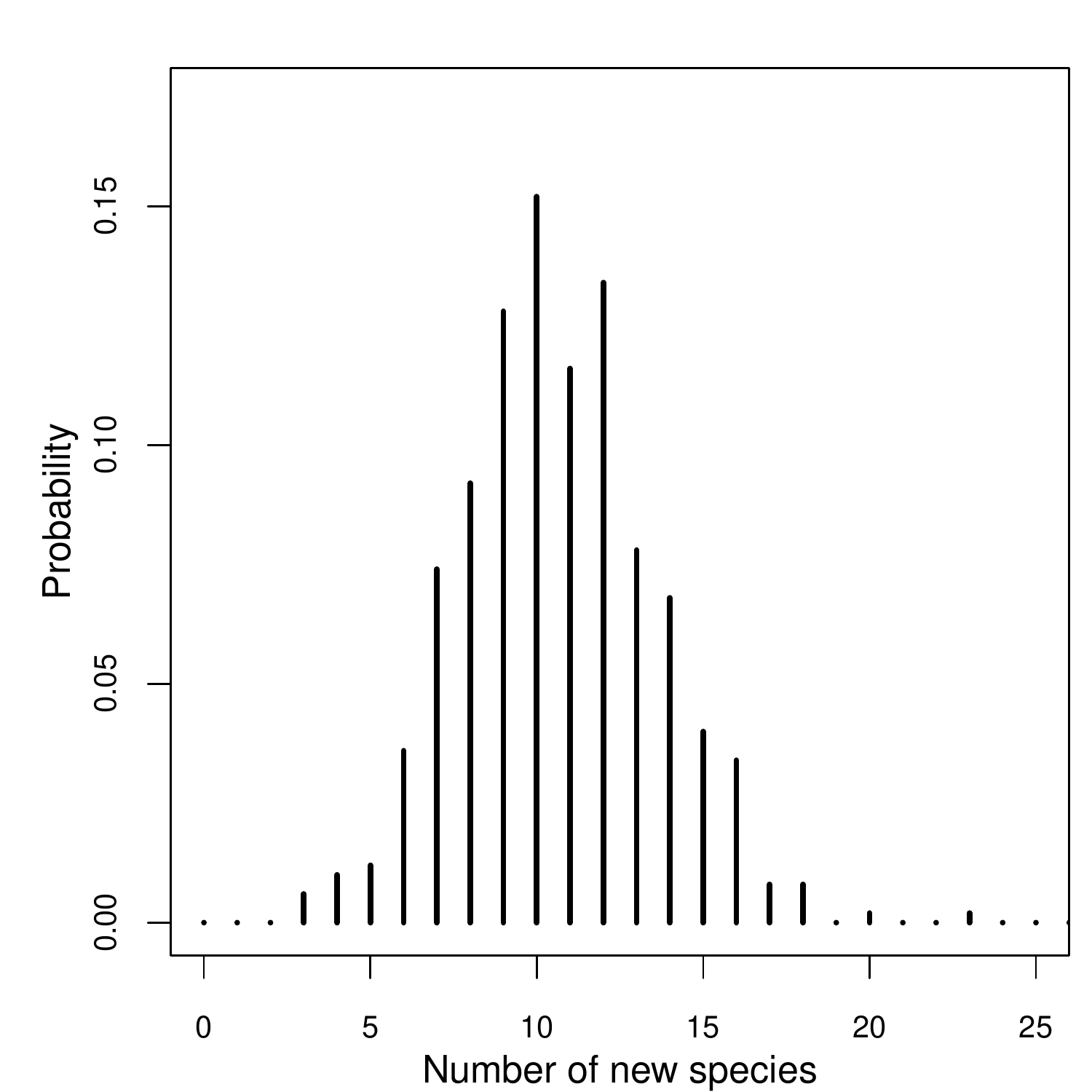}}
        \caption{Posterior distribution of the number of new species in a new sample of 100 individuals using purely simulations from the SSM and the formulas discussed in \citet{lijoi2007bayesian}. }\label{fig_CRP_NEW_SPECIES}
   \end{figure}

Once we implement these algorithms, we basically compute the empirical distribution of the simulated quantities to get the distribution of interest. We present our findings for the CRP. In this case $\tev=\te$, and the corresponding functions are given by
$$
p(\mv^n|\te) = \te^{K^n}\frac{\Ga(\te)}{\Ga(\te+n)}\,\prod_{j=1}^{K^n}\Ga(m_j^n)\,\,\,\text{and}\,\,\, q^n_k(\mv^n)=\left\{
                                                                                                                        \begin{array}{ll}
                                                                                                                          \frac{m_j^n}{\te+n}, & \hbox{$k\leq K^n$;} \\
                                                                                                                          \frac{\te}{\te+n}, & \hbox{$k=K^n+1$.}
                                                                                                                        \end{array}
                                                                                                                      \right.
$$
In addition, the the posterior distribution of the number of distinct species to be observed in the enlarged sample of size $n^*$ is given by
$$
\pr{K_{n^*}=k|\te, \mv^n} = \te^k\frac{(\te)_n}{(\te)_{n+n^*}} \sum_{l=k}^{n*} \binom{m}{l} |s(l,k)|\, (n)_{m_l} \,,\quad k=0,1,\ldots,n^*\,,
$$
where $(a)_n=a(a+1)\ldots(a+n-1)$ and $|s(n,k)|$ stands for the sign-less Stirling number of the first kind. Figure \ref{fig_CRP_NEW_SPECIES} displays the posterior distribution of the number of new species in a new sample of $n*=50$ individuals using these two approaches. Furthermore, Table \ref{tab_CRP_POST_new_species} shows posterior summaries for these distributions. Notice that in average both methods predict almost the same number of new species to be discovered in a new sample of size 50, although the simulation based approach prediction is a bit higher.

\begin{table}[ht]
\centering
\begin{tabular}{lccccc}
  \hline
 Method & Mean & Median & Sd & Q\,2.5\% & Q\,97.5\% \\
  \hline
  SSM based & 11.17 & 11.00 & 2.94 & 6.00 & 17.00 \\
  Lijoi's based & 10.62 & 10.00 & 2.90 & 5.00 & 16.00 \\
   \hline
\end{tabular}\caption{\footnotesize{Posterior summaries for the number of distinct species to be observed in the enlarged sample of size $n^*$ $K_{n^*}$. Computations based on an extension of the MCMC carried out to perform posterior inference about the parameter $\te$ of a CRP.}}\label{tab_CRP_POST_new_species}
\end{table}

%%%%%%%%%%%%%%%%%%%%%%%%%%%%%%%%%%%%%%%%%%%%%%%%%%%%%%%%%%%%%%%%%%%%%%%%%%%%%%%%%%%%%%%%%%%%%%%%%%%%%%%%%%%%%%%%%%%%
\section{Discussion}\label{sec_discussion}

In this work we have offered four detailed case studies that use Bayesian modeling techniques in order to answer complex questions. At every instance, we have provided specifics about model formulation, prior elicitation, simulation-based algorithms computation, and posterior inference. We hope that the reader benefits from such effort and considers the Bayesian paradigm a natural way to solve data-driven problems.

Even though the Bayesian approach to statistical inference is extremely beneficial, some challenges remain. In particular, we recommend consider alternative inference methods in order to account for ``big data'', which is currently an active research area in computational statistics (e.g., variational methods). See for example \cite{ormerod2010explaining}.

%\nocite{*}
\bibliography{references}

\begin{thebibliography}{}

\bibitem[Banerjee et~al., 2014]{banerjee2014hierarchical}
Banerjee, S., Carlin, B., and Gelfand, A. (2014).
\newblock {\em Hierarchical modeling and analysis for spatial data}.
\newblock Crc Press.

\bibitem[Gelman et~al., 2014]{gelman2014bayesian}
Gelman, A., Carlin, J., Stern, H., and Rubin, D. (2014).
\newblock {\em Bayesian data analysis}, volume~2.
\newblock Taylor \& Francis.

\bibitem[Jackman, 2009]{jackman2009bayesian}
Jackman, S. (2009).
\newblock {\em Bayesian analysis for the social sciences}, volume 846.
\newblock John Wiley \& Sons.

\bibitem[Lijoi et~al., 2007]{lijoi2007bayesian}
Lijoi, A., Mena, R.~H., and Pr{\"u}nster, I. (2007).
\newblock Bayesian nonparametric estimation of the probability of discovering
  new species.
\newblock {\em Biometrika}, 94(4):769--786.

\bibitem[M{\"u}ller et~al., 2015]{muller2015bayesian}
M{\"u}ller, P., Quintana, F.~A., Jara, A., and Hanson, T. (2015).
\newblock {\em Bayesian nonparametric data analysis}.
\newblock Springer.

\bibitem[M{\"u}ller and Rodriguez, 2013]{muller2013nonparametric}
M{\"u}ller, P. and Rodriguez, A. (2013).
\newblock {\em Nonparametric Bayesian Inference}.
\newblock NSF-CBMS regional conference series in probability and statistics.
  Inst Of Mathematical Stat.

\bibitem[Ormerod and Wand, 2010]{ormerod2010explaining}
Ormerod, J. and Wand, M.~P. (2010).
\newblock Explaining variational approximations.
\newblock {\em The American Statistician}, 64(2):140--153.

\bibitem[Pitman, 1995]{pitman1995exchangeable}
Pitman, J. (1995).
\newblock Exchangeable and partially exchangeable random partitions.
\newblock {\em Probability theory and related fields}, 102(2):145--158.

\bibitem[Pitman, 1996]{pitman1996some}
Pitman, J. (1996).
\newblock Some developments of the blackwell-macqueen urn scheme.
\newblock {\em Lecture Notes-Monograph Series}, pages 245--267.

\bibitem[Prado and West, 2010]{prado-10}
Prado, R. and West, M. (2010).
\newblock {\em Time Series: Modeling, Computation, and Inference}.
\newblock Chapman and Hall/CRC.

\bibitem[Robert, 2007]{robert2007bayesian}
Robert, C. (2007).
\newblock {\em The Bayesian choice: from decision-theoretic foundations to
  computational implementation}, volume~2.
\newblock Springer.

\bibitem[Rodr{\'\i}guez and Quintana, 2015]{rodriguez2015species}
Rodr{\'\i}guez, A. and Quintana, F.~A. (2015).
\newblock On species sampling sequences induced by residual allocation models.
\newblock {\em Journal of statistical planning and inference}, 157:108--120.

\bibitem[Taddy and Kottas, 2010]{taddy2010bayesian}
Taddy, M. and Kottas, A. (2010).
\newblock A bayesian nonparametric approach to inference for quantile
  regression.
\newblock {\em Journal of Business \& Economic Statistics}, 28(3):357--369.

\bibitem[Tsay, 2010]{tsay2010analysis}
Tsay, R.~S. (2010).
\newblock {\em Analysis of Financial Time Series}.
\newblock CourseSmart. Wiley.

\bibitem[West and Harrison, 1999]{west1999bayesian}
West, M. and Harrison, J. (1999).
\newblock {\em Bayesian Forecasting and Dynamic Models}.
\newblock Springer Series in Statistics. Springer New York.

\end{thebibliography}
\bibliographystyle{apalike}

\appendix

\section{Notation}

The cardinality of a set $A$ is denoted by $|A|$. If P is a logical proposition, then $1_{\text{P}} = 1$ if P is true, and $1_{\text{P}} = 0$ if P is false. $\floor{x}$ denotes the floor of $x$, whereas $[n]$ denotes the set of all integers from 1 to $n$, i.e., $\{1,\ldots,n\}$. The Gamma function is given by $\Gamma(x)=\int_0^\infty u^{x-1}\,e^{-u}\,\text{d}u$. 

Matrices and vectors with entries consisting of subscripted variables are denoted by a boldfaced version of the letter for that variable. For example, $\xv = (x_1,\ldots, x_n)$ denotes an $n\times1$ column vector with entries $x_1,\ldots, x_n$. We use $\zerov$ and $\boldsymbol{1}$ to denote the column vector with all entries equal to 0 and 1, respectively, and $\Ima$ to denote the identity matrix. A subindex in this context refers to the corresponding dimension; for instance, $\Ima_n$ denotes the $n\times n$ identity matrix. The transpose of a vector $\xv$ is denoted by $\xv^\trans$; analogously for matrices. Moreover, if $\Xm$ is a square matrix, we use $\text{tr}(\Xm)$ to denote its trace and $\Xm^{-1}$ to denote its inverse. The norm of $\xv$, given by $\sqrt{\xv^\trans\xv}$, is denoted by $\|\,\xv\|\,$.

\end{document}